\documentclass[journal,onecolumn,draftclsnofoot,]{IEEEtran}
\usepackage{setspace}
\usepackage{enumerate}
\usepackage[dvipsnames]{xcolor}
\usepackage{graphicx}
\usepackage{epstopdf}
\usepackage{csquotes}
\usepackage{subfig}
\usepackage{mathtools}
\usepackage{amssymb, amsmath,amsthm, amsfonts}
\usepackage{ifthen}
\usepackage{tikz}
\usepackage{cite}
\usepackage{enumerate}
\usepackage{verbatim}
\usepackage{pifont}
\usepackage{bm}  
\usepackage{stackengine}
\usepackage{bbm}

\usepackage{accents}

\DeclareMathOperator*{\argmax}{arg\,max}
\DeclareMathOperator*{\argmin}{arg\,min}
\DeclareSymbolFont{bbold}{U}{bbold}{m}{n}
\DeclareSymbolFontAlphabet{\mathbbold}{bbold}

\newtheorem{theorem}{Theorem}
\newtheorem{cor}[theorem]{Corollary}

\newtheorem{prop}[theorem]{Proposition}

\newtheorem{remark}[theorem]{Remark}

\newtheorem{definition}[theorem]{Definition}
\newtheorem{example}{Example}

\newcommand{\I}{\mathcal{I}}

\newcommand{\ind}{\mathbbm 1}
\newcommand{\Y}{\mathcal{Y}}
\newcommand{\Ucal}{\mathcal{U}}
\newcommand{\W}{\mathcal{W}}
\newcommand{\V}{\mathcal{V}}
\newcommand{\X}{\mathcal{X}}
\newcommand{\M}{\mathcal{M}}

\newcommand{\Scal}{\mathcal{S}}
\newcommand{\Ccal}{\mathcal{C}}

\newcommand\blfootnote[1]{%
	\begingroup
	\renewcommand\thefootnote{}\footnote{#1}%
	\addtocounter{footnote}{-1}%
	\endgroup
}

\def\h2{\tilde h}

\def\hm1{\hat h_{-1}}

\begin{document}
\title{ Distributed Hypothesis Testing Over Discrete Memoryless Channels}
\author{Sreejith Sreekumar and Deniz G\"und\"uz \\ Imperial College London, UK \\
Email: \{s.sreekumar15, d.gunduz\}@imperial.ac.uk}

\maketitle

\begin{abstract}
A distributed binary hypothesis testing (HT) problem involving two parties, one referred to as the observer and the other as the detector is studied. The observer observes a discrete memoryless source (DMS) and  communicates its observations to the detector over a discrete memoryless channel (DMC). The detector observes another DMS correlated with that at the observer, and  performs a binary HT on the joint distribution of the two DMS's using its own observed data and the information received from the observer. The trade-off between the type I error probability and the type II error-exponent of the HT is explored.
Single-letter lower bounds on the optimal  type II error-exponent are obtained  by using two different coding schemes, a separate HT and channel coding scheme and a joint HT and channel coding scheme based on hybrid coding for the matched bandwidth case. Exact single-letter characterization of the same is established for the special case of testing against conditional independence, and it is shown to be achieved by the separate HT and channel coding scheme. An example is provided where  the joint scheme achieves a strictly better performance than the separation based scheme.
\end{abstract}

\section{Introduction}
\begin{figure}[t]
\centering
\includegraphics[trim=0.6cm 7cm 0cm 7cm, clip, width= 0.6\textwidth]{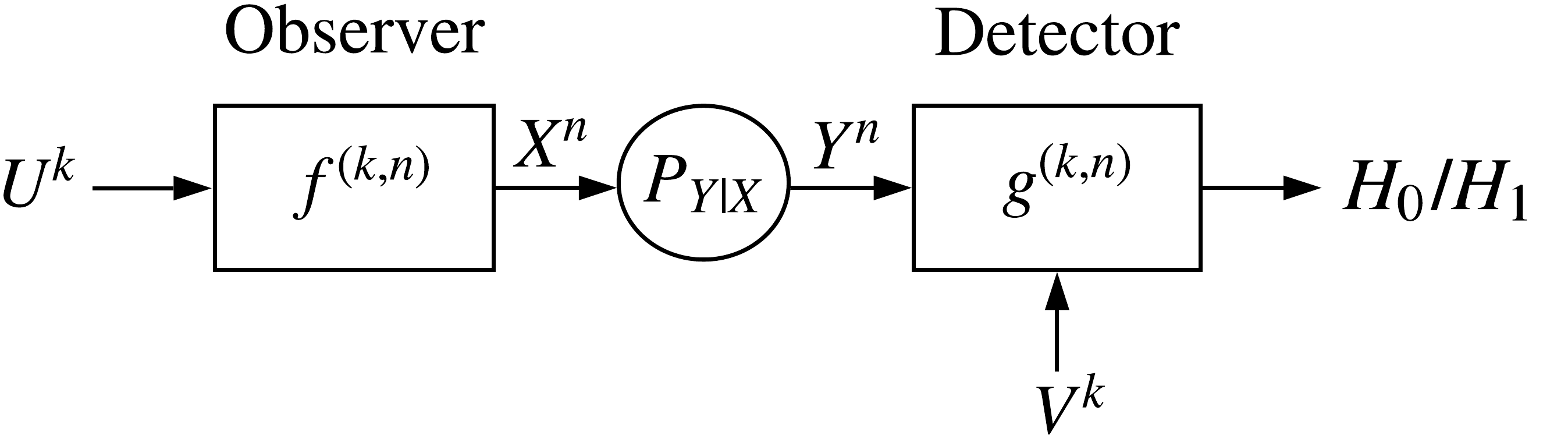}
\caption{Distributed HT over a DMC.} \label{htnoisymodel}
\end{figure}

\blfootnote{This work is supported in part by the European Research Council (ERC) through Starting Grant BEACON (agreement \#677854). A part of this work was presented at the International Symposium on Information theory (ISIT), Aachen, 2017 \cite{Sree_isit17}.}
Given data samples, statistical hypothesis testing (HT) deals with the problem of ascertaining the true assumption, that is, the true hypothesis, about the data from among a set of hypotheses. In modern communication networks (like in sensor networks, cloud computing and Internet of things (IoT)), data  is gathered at multiple remote nodes, referred to as  \textit{observers}, and transmitted over noisy links to another node for further processing. Often, there is some prior statistical knowledge available about the data, for example, that the joint probability distribution of the data belongs to a certain prescribed set. In such scenarios, it is of interest to identify the true underlying probability distribution, and this naturally leads to the problem of distributed HT over noisy channels. The simplest case of such a scenario is depicted in Fig. \ref{htnoisymodel}, where there is a  single observer and two possibilities for the joint distribution of the data. The observer observes $k$ independent and identically distributed (i.i.d) data samples $U^k$, and communicates its observation to the detector by $n$ uses of the DMC, characterized by the conditional distribution $P_{Y|X}$. The detector performs a binary hypothesis test on the joint distribution of the data $(U^k,V^k)$ to decide between them, based on the channel outputs $Y^n$ as well as its own observations $V^k$. The null and the alternate hypothesis of the hypothesis test are given by 
\begin{subequations}\label{HTdef}
\begin{equation} 
  H_0: (U^k,V^k) \sim \prod_{i=1}^k P_{UV},  
\end{equation}
and
\begin{equation}
H_1: (U^k,V^k) \sim \prod_{i=1}^k Q_{U V},   
\end{equation}
\end{subequations}
 respectively. Our goal is to characterize the optimal exponential rate of decay of the type II error probability asymptotically, known as the \textit{ type II  error-exponent (henceforth, also referred to as error-exponent)} for a prescribed constraint on  the type I error probability for the above hypothesis test. 

 In the centralized scenario, in which the detector performs a binary hypothesis test on the probability distribution of the data it observes directly, the optimal error-exponent is characterized by the well-known lemma of Stein \cite{Chernoff-1952} (see also \cite{Hoeff-1965}).
The study of distributed statistical inference under communication constraints was conceived by Berger in \cite{Berger_1979}. In \cite{Berger_1979}, and in the follow up literature summarized below, communication from the observers to the detector are assumed to be over rate-limited error-free channel. Some of the fundamental results in this setting for the case of a single observer was established by Ahlswede and Csisz\'{a}r in \cite{Ahlswede-Csiszar}. They obtained a tight single-letter characterization  of the optimal error-exponent for a special case of HT known as \textit{testing against independence} (TAI), in which, $Q_{UV}=P_{U} \times P_V$. Furthermore, the authors established a lower bound on the optimal error-exponent for the general HT case, and proved a \textit{strong converse} result, which states that the optimal achievable error-exponent is independent of the constraint on the type I error probability. A tighter lower bound for the general HT problem  is established by Han \cite{Han}, which recovers the corresponding lower bound in \cite{Ahlswede-Csiszar}. Han also considered complete data compression in a related setting where either $U$, or $V$, or both (also referred to as two-sided compression setting) are compressed and communicated to the detector using a message set of size two. It is shown that, asymptotically, the optimal error-exponent achieved in these three settings are equal. In contrast, a single-letter characterization of the optimal error-exponent for even the TAI with two-sided compression and general rate constraints remains open till date. Shalaby et al. \cite{Shalaby-pap} extended the complete data compression result of Han to show that the optimal error-exponent is not improved even if the rate constraint is relaxed to that of zero-rate compression (sub-exponential message set with respect to blocklength $k$). Shimokawa et al. \cite{Shimokawa} obtained a tighter lower bound on the optimal error-exponent for general HT by considering quantization and binning at the encoder along with a minimum empirical-entropy decoder. Rahman and Wagner \cite{Rahman-Wagner} studied the setting with multiple observers, in which, they showed that for the case of a single-observer, the \textit{quantize-bin-test} scheme achieves the optimal error-exponent for \textit{testing against conditional independence} (TACI), in which, $V=(E,Z)$ and $Q_{UEZ}=P_{UZ}P_{E|Z}$. Extensions of the distributed HT problem has also been considered in several other interesting scenarios involving multiple detectors \cite{Wigger-Timo}, multiple observers \cite{Zhao-Lai}, interactive HT \cite{Xiang-Kim-1, Xiang-Kim-2}, collaborative HT \cite{Katz-collab}, HT with lossy source reconstruction \cite{Katz-estdetjourn}, HT over a multi-hop relay network \cite{Sadaf-Wigger-Li}, etc., in which, the authors obtain a single-letter characterization of the optimal error-exponent in some special cases.

While the works mentioned above have studied the unsymmetric case of focusing on the error-exponent for a constraint on the type I error probability, other works have analyzed the trade-off between the type I and type II error probabilities in the exponential sense. In this direction, the optimal trade-off between the type I and type II error-exponents in the centralized scenario is obtained in \cite{Blahut-1974}. The distributed version of this problem is first studied in  \cite{HK-1989}, where inner bounds on the above trade-off are established. This problem has also been explored from an information-geometric perspective for the zero-rate compression scenario in \cite{AH-1989} and \cite{Han-Amari-1989}, which provide further insights into the geometric properties of the optimal trade-off between the two exponents. A Neyman-Pearson like test  in the zero-rate compression scenario is proposed in \cite{Watanabe-2017}, which, in addition to achieving the optimal trade-off between the two exponents, also achieves the optimal second order asymptotic performance among all symmetric (type-based) encoding schemes. However, the optimal trade-off between the type I and type II error-exponents for the general distributed HT problem remains open. Recently, an inner bound for this trade-off is obtained in \cite{WK-2017}, by using the reliability function of the optimal channel detection codes.

In contrast, HT in distributed settings that  involve communication over noisy channels has not been considered until now. In noiseless rate-limited settings, the encoder can reliably communicate its observation subject to a rate constraint. However, this is no longer the case in noisy settings, which complicates the study of error-exponents in HT. Since the capacity of the channel $P_{Y|X}$, denoted by $C(P_{Y|X})$, quantifies the maximum rate of reliable communication over the channel, it is reasonable to expect that it plays a role in the characterization of the optimal error-exponent similar to the rate-constraint $R$ in the noiseless setting. Another measure of the noisiness of the channel is the so-called \textit{reliability function} $E(R,P_{Y|X})$ \cite{Csiszar-Korner}, which is defined as the maximum achievable exponential decay rate of the probability of error (asymptotically) with respect to the blocklength for message rate of $R$. It appears natural that the reliability function plays a role in the characterization of the achievable error-exponent for distributed HT over a noisy channel. Indeed, in Theorem \ref{thm:shtcceed} given below, we provide a lower bound on the optimal error-exponent that depends on the \textit{expurgated exponent} at rate $R$, $E_x(R,P_{Y|X})$, which is a lower bound on $E(R,P_{Y|X})$\cite{Gallager-codthm}. However, surprisingly, it will turn out that the reliability function does not play a role in the characterization of the error-exponent for TACI in the regime of vanishing type I error probability constraint.

The goal of this paper is to study the best attainable error-exponent for distributed HT over a DMC with a single observer and obtain a computable characterization of the same. Although a complete solution is not to be expected for this problem (since even the corresponding noiseless case is still open), the aim is to provide an achievable scheme for the general problem, and to identify special cases in which a tight characterization can be obtained. 
In the sequel, we first introduce
a separation based scheme that performs independent hypothesis testing and channel coding, which we refer to as the \textit{separate hypothesis testing and channel coding} (SHTCC) scheme. This scheme combines the Shimokawa-Han-Amari scheme \cite{Shimokawa}, which is the best known coding scheme till date for distributed HT over a rate-limited noiseless channel, with the channel coding scheme that achieves the expurgated exponent \cite{Gallager-codthm}\cite{Csiszar-Korner} of the channel along with the best channel coding error-exponent for a single special message. The channel coding scheme is based on the Borade-Nakibo\u{g}lu-Zheng unequal error-protection scheme \cite{Borade-09}. As we show later, the SHTCC scheme achieves the optimal error-exponent for TACI.

Although the SHTCC scheme is attractive due to  its modular design, \textit{joint source channel coding} (JSCC) schemes are known to outperform separation based schemes in several different contexts, for example, the error exponent for reliable transmission of a source over a DMC \cite{Csiszar-1980}, reliable transmission of correlated sources over a multiple-access channel \cite{Cover-elgamal-salehi}, etc., to name a few. While in separation based schemes coding is usually performed by first quantizing the observed source sequence to an index, and transmitting the channel codeword corresponding to that index (independent of the source sequence), JSCC schemes allow the channel codeword to be dependent on the source sequence, in addition to the quantization index. Motivated by this, we propose a second scheme, referred to as the \textit{joint HT and channel coding} (JHTCC) scheme, based on \textit{hybrid coding} \cite{Lim-minero-kim-2015} for the communication between the observer and the detector. 

 Our main contributions can be summarized as follows. 
\begin{enumerate}[(i)]
    \item We propose two different coding schemes (namely, SHTCC and JHTCC) for distributed HT over a DMC, and analyze the error-exponents achieved by these schemes.
    \item We obtain an exact single-letter characterization of the optimal error-exponent for the special case of TACI with a vanishing type I error probability constraint, and show that it is achievable by the SHTCC scheme.
    \item We provide an example where the JHTCC scheme achieves a strictly better error-exponent than the SHTCC scheme.
\end{enumerate}

The rest of the paper is organized as follows. In Section \ref{Prelims}, we introduce the notations, detailed system model and definitions. Following this, we introduce the main results in Section \ref{results-main} and \ref{Optresults}. The achievable schemes are presented in Section \ref{results-main} and the optimality results for special cases are discussed in Section \ref{Optresults}. Finally, Section \ref{sec:conclu} concludes the paper.

\section{Preliminaries} \label{Prelims}
\subsection{Notations}  
Random variables (r.v.'s) are denoted by capital letters (e.g., $X$), their realizations by the corresponding lower case letters (e.g., $x$), and their support by calligraphic letters (e.g., $\X$). The cardinality of  a finite set $\X$ is denoted by $|\X|$. The set of all probability distributions on alphabet $\X$ is denoted by $\mathcal{P}_{\X}$.  Similar notations apply for set of conditional probability distributions, e.g., $\mathcal{P}_{\Y|\X}$.
$X-Y-Z$ denotes that $X,~Y$ and $Z$ form a Markov  chain. 
For $m \in \mathbb{Z}^+$, $X^m$ denotes the sequence $X_1, \ldots, X_m$.
 Following the notation in \cite{Csiszar-Korner}, for a probability distribution $P_X$ on r.v. $X$, $T_{P_X}^m$ and $T_{[P_X]_{\delta}}^m$ (or $T_{[X]_{\delta}}^m$) denote the set of sequences $x^m \in \X^m$ of type $P_X$  and the set of $P_X$-typical sequences, respectively. The set of all possible types of sequences of length $m$ with alphabet $\mathcal{X}$ is denoted by $\mathcal{T}^m_{\X}$, and $\cup_{m \in \mathbb{Z}^+}\mathcal{T}^m_{\X} $ is denoted by $\mathcal{T}_{\X}$. Similar notations apply for pair's and other larger combinations of r.v.'s, e.g., $T_{P_{XY}}^m$ $T_{[P_{XY}]_{\delta}}^m$, $\mathcal{T}^m_{\X \Y}$, $\mathcal{T}_{\X\Y}$, etc.. The standard information theoretic quantities like Kullback-Leibler (KL) divergence between distributions $P_X$ and $Q_X$, the entropy of $X$ with distribution $P_X$, the conditional entropy of $X$ given $Y$ and the mutual information  between  $X$ and $Y$ with joint distribution $P_{XY}$, are denoted by
 $D(P_X||Q_X)$,  $H_{P_X}(X)$, $H_{P_{XY}}(X|Y)$ and $I_{P_{XY}}(X;Y)$, respectively. 
 When the distribution of the r.v.'s involved are clear from the context, the last three quantities are denoted simply by $H(X)$, $H(X|Y)$ and $I(X;Y)$, respectively.
 Given  realizations $X^m=x^m$ and $Y^m=y^m$, $H_e(x^m|y^m)$ denotes the conditional empirical entropy defined as
 \begin{align}
 H_e(x^m|y^m):= H_{P_{\tilde X \tilde Y}}(\tilde X|\tilde Y),
 \end{align}
 where $P_{\tilde X \tilde Y}$ denote the joint type of $(x^m,y^m)$, and $:=$ represents equality by definition (throughout this paper). 
      For $a \in \mathbb{R}^+$, $[a]$ denotes the set of integers $\{1, 2,\ldots, \lceil a \rceil \}$.
All logarithms considered in this paper are with respect to the base $e$ unless specified otherwise. For any set $\mathcal{G}$, $\mathcal{G}^c$ denotes the set complement. 
  $a_k \xrightarrow{(k)} b$ represents  $\lim_{k \rightarrow \infty} a_k=b$. Similar notations are used for inequalities that hold asymptotically, e.g., , $a_k\overset{(k)}{\geq } b_k$ denotes  $\lim_{k \rightarrow \infty}a_k \geq b$. $\mathbb{P}(\mathcal{E})$ denotes the probability of event $\mathcal{E}$. For functions $f_1:\mathcal{A} \rightarrow \mathcal{B}$ and  $f_2: \mathcal{B} \rightarrow \mathcal{C} $, $f_2 \circ f_1$ denotes function composition. Finally, $\ind(\cdot)$ denotes the indicator function, and $O(\cdot)$ and $o(\cdot)$ denote the standard asymptotic notation.
\subsection{Problem formulation} \label{probformulanddef}
All the r.v.'s considered henceforth are discrete with finite support. Unless specified otherwise, we will denote the probability distribution of a r.v. $Z$ under the null and alternate hypothesis by $P_Z$ and $Q_Z$, respectively. Let $k,n \in \mathbb{Z}^+$ be arbitrary. The encoder (at the observer) observes $U^k$,  and transmits codeword $X^n= f^{(k,n)}(U^k)$, where $f^{(k,n)}: \Ucal^k \rightarrow \X^n$ represents the encoding function (possibly stochastic).
Let  $\tau := \frac{n}{k}$ denote the \textit{bandwidth ratio}. 
The channel  output $Y^n$ is given by  the probability law
\begin{align}
    P_{Y^n| X^n }(y^n|x^n)= \prod_{j=1}^n P_{Y|X}(y_{j}|x_{j}), \label{DMCprbdist}
\end{align}
i.e., the channels between the  observers  and  the detector are independent of each other and memoryless. 
Depending on the received symbols  $Y^n$ and its own observations $V^k$, the detector makes a decision  between the two hypotheses $H_0$ and $H_1$ given in \eqref{HTdef}. Let $H \in \{0,1\}$ denote the actual hypothesis and $\hat H \in \{0,1\}$  denote the output of the hypothesis test, where $0$ and $1$ denote $H_0$ and $H_1$, respectively, and $\mathcal{A}_{(k,n)} \subseteq \Y^n \times \V^k $ denote the acceptance region for $H_0$. Then, the decision rule $g^{(k,n)}:\Y^n \times \V^k \rightarrow \{0,1\}$  is given by
 \begin{align*}
g^{(k,n)}\left(y^n,v^k\right)= 1-\ind \left(\left(y^n,v^k\right) \in \mathcal{A}_{(k,n)} \right).
 \end{align*} 
Let  
\begin{align}
     \alpha \left(k,n, f^{(k,n)}, g^{(k,n)} \right)&:=1- P_{Y^{n} V^{k}}\left(\mathcal{A}_{(k,n)}\right), \notag \\
   \mbox{and } \beta \left(k,n, f^{(k,n)}, g^{(k,n)} \right) &:= Q_{Y^{n} V^{k}}\left(\mathcal{A}_{(k,n)}\right), \notag
\end{align}
 denote the type I and type II error probabilities for the encoding function $f^{(k,n)}$ and decision rule $g^{(k,n)}$, respectively.
\begin{definition} \label{defexpdistach} 
An error-exponent $\kappa$ is $(\tau, \epsilon)$ achievable if there exists  a sequence of  integers $k$, corresponding sequences of encoding function $f^{(k,n_k)}$ and decision rules $g^{(k,n_k)}$  such that $n_k \leq \tau k$, $\forall~k$, 
\begin{subequations} \label{seqkappa}
\begin{equation}  
  \liminf_{k \rightarrow \infty}  \frac{-1}{k}\log \left( \beta \left(k,n_k, f^{(k,n_k)}, g^{(k,n_k)} \right) \right)  \geq \kappa,  
\end{equation}
\begin{equation} \label{t1errprbconst}
   \mbox{and } \limsup_{k \rightarrow \infty} \alpha\left(k,n_k, f^{(k,n_k)}, g^{(k,n_k)} \right) \leq \epsilon.  
\end{equation}
\end{subequations} 
\end{definition}
For $(\tau, \epsilon) \in \mathbb{R}^+ \times [0,1]$, let
\begin{align} 
\kappa (\tau, \epsilon) & :=  \sup \{ \kappa': \kappa' \mbox{ is } (\tau,\epsilon) \mbox{ achievable} \}. \label{opterror-exponentdefsup}
 \end{align}
We are interested in obtaining a computable characterization of $\kappa (\tau, \epsilon)$.

It is well known that the Neyman-Pearson test \cite{NP-1933} gives the optimal trade-off between the type I and type II error probabilities, and hence, also between the error-exponents in HT. It follows that the optimal error-exponent for distributed HT over a DMC is achieved when the channel-input $X^n$ is generated correlated with $U^k$ according to some optimal conditional distribution $P_{X^n|U^k}$, and the optimal Neyman-Pearson test 
is performed on the data available (both received and observed) at the detector. It can be shown, similarly to \cite[Theorem 1]{Ahlswede-Csiszar}, that the optimal  error-exponent for vanishing type I error probability constraint is characterized by the multi-letter expression (see \cite{SD_isit19}) given by
\begin{align}
\lim_{\epsilon \rightarrow 0}\kappa(\tau,\epsilon)=   \sup_{\substack{P_{X^n|U^k} \in~ \mathcal{P}_{\X^n|\Ucal^k}, \\ k, n ~\in ~\mathbb{Z}^+,~n \leq \tau k}}\frac{1}{k}D\left(P_{Y^nV^k}||Q_{Y^nV^k}\right). \label{multletexpgen}
\end{align}
However, the above expression does not single-letterize in general, and hence, is intractable as it involves optimization over large dimensional probability simplexes when $k$ and $n$ are large. Moreover, the encoder and the detector of a  scheme achieving the error-exponent given in \eqref{multletexpgen}  would be computationally complex to implement from a practical viewpoint. Consequently, we establish two computable single-letter lower bounds on $\kappa(\tau, \epsilon)$ in the next section by using the SHTCC and JHTCC schemes.
\section{Achievable schemes} \label{results-main}
In \cite{Shimokawa}, Shimokawa et al. obtained a lower bound on the optimal error-exponent for distributed HT over a rate-limited noiseless channel by using a coding scheme that involves quantization and binning at the encoder. In this scheme, the type\footnote{Since the number of types is polynomial in the blocklength, these can be communicated error-free at asymptotically zero-rate.} of the observed sequence $U^k=u^k$ is transmitted by the encoder to the detector, which is useful to improve the performance of the hypothesis test. In fact, in order to achieve the error-exponent proposed in \cite{Shimokawa}, it is sufficient to send a message 
indicating whether $U^k$ is typical or not, rather than sending the exact type of $U^k$. Although it is not possible to get perfect reliability for messages transmitted over a noisy channel, intuitively, it is desirable to protect the typicality information about the observed sequence as reliably as possible. Based on this intuition, we next propose the SHTCC scheme that performs independent HT and channel coding and protects the message indicating whether $U^k$ is typical or not, as reliably as possible. 
\subsection{SHTCC Scheme:} \label{ssshtcc}
 In the SHTCC scheme, the encoding and decoding  functions are restricted to be of the form  $f^{(k,n)}= f_c^{(k,n)} \circ f_s^{(k)}$  and $g^{(k,n)}=   g_s^{(k)} \circ g_c^{(k,n)}$, respectively. The source encoder $f_s^{(k)}: \Ucal^k \rightarrow \M=\{0,1, \cdots, \lceil e^{kR} \rceil\}$ generates an index $M= f_s^{(k)}(U^k)$ and the channel encoder $f_c^{(k,n)}: \M \rightarrow \tilde{\Ccal}=\{X^n(j),~j \in [0:\lceil e^{kR} \rceil] \}$ generates the channel-input codeword $X^n=f_c^{(k,n)}(M)$. Note that the rate of this coding scheme is $\frac{kR}{n}= \frac{R}{\tau}$ bits per channel use.
The channel decoder $g_c^{(k,n)}: \Y^n \rightarrow \M$ maps the channel-output $Y^n$ into an index $\hat M=g_c^{(k,n)}(Y^n)$,  and $g_s^{(k)}: \M \times \V^k \rightarrow \{0,1\}$ outputs the result of the HT as $\hat H=g_s^{(k)}(\hat M,V^k)$.  Note that $f_c^{(k,n)}$ depends on $U^k$ only through the output of $f_s^{(k)}(U^k)$ and $g_c^{(k,n)}$ depends on $V^k$ only through $Y^n$. Hence, the scheme is modular in the sense that $(f_c^{(k,n)},g_c^{(k,n)})$ can be designed independent of $(f_s^{(k)},g_s^{(k)})$. In other words, any good channel coding scheme  may be used in conjunction with a good compression scheme. If $U^k$ is not typical according to $P_U$, $f_s^{(k)}$ outputs a \textit{special} message, referred to as the \textit{error} message, denoted by $M=0$, to inform the detector to declare $\hat H=1$. There is obviously a trade-off between the reliability of the error message and the other messages in channel coding. The best known reliability for protecting a single \textit{special} message when the other messages $M \in [e^{nR}]$ of rate $R$, referred to as \textit{ordinary} messages, are required to be communicated reliably is given by the \textit{red-alert exponent} in \cite{Borade-09}. The red-alert exponent is defined as
\begin{align}
E_m(R,P_{Y|X}):= \max_{\substack{P_{SX}:~ \mathcal{S}=\mathcal{X},\\I(X;Y|S)=R,\\S-X-Y} }\sum_{s \in \mathcal{S}}P_{S}(s)~D\left(P_{Y|S=s}||P_{Y|X=s}\right).    
\end{align}
Borade et al.'s  scheme uses an appropriately generated codebook along with a two-stage decoding procedure. The first stage is a \textit{joint-typicality} decoder to decide whether $X^n(0)$ is transmitted, while the second stage is a \textit{maximum-likelihood decoder} to decode the ordinary message if the output of the first stage is not zero, i.e., $\hat M \neq 0$. 
On the other hand, it is well-known that if the rate of the messages is $R$, a channel coding error-exponent equal to $E_x(R,P_{Y|X})$ is achievable, where
\begin{align}
 & E_x(R,P_{Y|X}):= \max_{P_X} \max_{\rho \geq 1} \left\lbrace -\rho ~R-\rho~\log \left( \sum_{x,\tilde x}P_{X}(x)P_{X}(\tilde x) \left( \sum_{y} \sqrt{P_{Y|X}(y|x)P_{Y|X}(y|\tilde x)}\right)^{\frac{1}{\rho}} \right)\right\rbrace, 
\end{align} 
is the \textit{expurgated} exponent at rate $R$ \cite{Gallager-codthm}\cite{Csiszar-Korner}. 
Let 
\begin{align}
E_m(P_{SX},P_{Y|X})&:= \sum_{s \in \mathcal{S}}P_{S}(s)~D\left(P_{Y|S=s}||P_{Y|X=s}\right),  
\end{align}
where, $\mathcal{S}=\X$ and $S-X-Y$, and 
\begin{align} 
 &E_x(R,P_{SX},P_{Y|X})& \notag \\
 &:= \max_{\rho \geq 1} \left\lbrace -\rho ~R-\rho~\log \left( \sum_{s,x,\tilde x}P_S(s)P_{X|S}(x|s)P_{X|S}(\tilde x|s) \left( \sum_{y} \sqrt{P_{Y|X}(y|x)P_{Y|X}(y|\tilde x)}\right)^{\frac{1}{\rho}} \right)\right\rbrace. \notag
\end{align} 
  Although Borade et al.'s scheme is concerned only with the reliability of the special message, it is not hard to see using the technique of \textit{random-coding} that for a fixed distribution $P_{SX}$, there exists a codebook $\tilde C$, and encoder and decoder as in Borade et al.'s  scheme, such that the rate is $0 \leq R \leq I(X;Y|S)$ and the special message achieves a reliability equal to $E_m(P_{SX},P_{Y|X})$, while the ordinary messages achieve a reliability equal to $E_x(R, P_{SX},P_{Y|X})$.
Note that $E_m(P_{SX},P_{Y|X})$ and $E_x(R,P_{SX},P_{Y|X})$ 
denote Borade et al.'s  red-alert exponent and the expurgated exponent with fixed distribution $P_{SX}$, respectively, and that both are  inter-dependent through $P_{SX}$. Thus, varying $P_{SX}$ provides a trade-off between the reliability for the ordinary messages and the special message. We will use Borade et al.'s  scheme for channel coding in the SHTCC scheme, such that the error message and the other messages correspond to the special and ordinary messages, respectively. The SHTCC scheme will be described in detail in Appendix \ref{SHTCCproof}. We next state a lower bound on $\kappa(\tau,\epsilon)$ that is achieved by the SHTCC scheme. For brevity, we will use the shorter notations $C$, $E_m(P_{SX})$ and $E_x(R,P_{SX})$ instead of $C(P_{Y|X})$, $E_m(P_{SX},P_{Y|X})$ and  $E_x(R,P_{SX},P_{Y|X})$, respectively. 
 \begin{theorem} \label{thm:shtcceed}
For $\tau \geq 0$, $\kappa(\tau, \epsilon) \geq \kappa_s(\tau)$, $ \forall ~\epsilon \in (0,1]$, where 
\begin{flalign}
&\kappa_s(\tau) & \notag \\
&:= \sup_{\substack{(P_{W|U},P_{SX},R) \\ \in~ \mathcal{B}(\tau, P_{Y|X})}} \min \left\lbrace E_1(P_{W|U}),~ E_2(P_{W|U},P_{SX},\tau),~E_3(P_{W|U},P_{SX},\tau),~E_4(P_{W|U},P_{SX},\tau) \right\rbrace, \label{error-exponentshtcc} &&
\end{flalign}
where
\begin{flalign} \label{eqsetsepvalid}
\mathcal{B}\left( \tau, P_{Y|X}\right):=  \left\{
  \begin{aligned}
       (P_{W|U},P_{SX},R):~&\mathcal{S}=\X,~P_{UVWSXY}(P_{W|U},P_{SX}):= P_{UV}P_{W|U}P_{SX}P_{Y|X},\\&~I_{P}(U;W|V) \leq R <  \tau I_P(X;Y|S)  
  \end{aligned}
\right\},&&
\end{flalign}
\begin{flalign}
&E_1(P_{W|U}):= \min_{P_{\tilde U \tilde V \tilde W} \in \mathcal{T}_1(P_{UW}, P_{VW}) } D(P_{\tilde U \tilde V \tilde W} ||  Q_{UVW}), \label{expterm1} &&
\end{flalign}
\begin{flalign} 
& E_2(P_{W|U},P_{SX},R) \notag \\
&:=
\begin{cases}  \label{expterm2}
\substack{ \min \\ P_{\tilde U \tilde V \tilde W} \in \mathcal{T}_2(P_{UW}, P_{V}) }~ D(P_{\tilde U \tilde V \tilde W} || Q_{UVW})+R -I_P(U;W|V), \qquad \mbox{if } I_P(U;W) > R, \\  \qquad  ~~ \qquad \qquad \qquad \qquad \qquad   \infty, \qquad  \qquad \qquad \qquad\qquad \qquad  \qquad \mbox{otherwise}, 
 \end{cases} &&
\end{flalign}
\begin{flalign} 
 &E_3(P_{W|U},P_{SX},R,\tau) \notag \\
 &:= 
\begin{cases} \label{chnerrcase2}
\substack{ \min \\ P_{\tilde U \tilde V \tilde W} \in \mathcal{T}_3(P_{UW}, P_{V}) }~  D(P_{\tilde U \tilde V \tilde W} || Q_{UVW}) +R-I_P(U;W|V) + ~\tau E_x \left(\frac{R}{\tau},P_{SX}\right), \mbox{if } I_P(U;W) > R, \\
\substack{ \min \\ P_{\tilde U \tilde V \tilde W} \in \mathcal{T}_3(P_{UW}, P_{V}) }~  D(P_{\tilde U \tilde V \tilde W} || Q_{UVW})+I_P(V;W) +\tau E_x\left(\frac{R}{\tau},P_{SX}\right),\qquad\qquad  \quad \mbox{otherwise}, 
\end{cases} \\
 &E_4( P_{W|U},P_{SX},R,\tau) 
 :=
\begin{cases}  \label{chnerrcase3}
D(P_{ V} || Q_{V})+R-I_P(U;W|V)+ \tau E_m\left(P_{SX}\right), \qquad \quad \mbox{if } I_P(U;W) > R, \\
  D(P_{V } || Q_{V})+I_P(V;W) + \tau E_m\left(P_{SX}\right),\qquad \qquad \qquad \qquad \mbox{otherwise,} 
 \end{cases} \\
 &Q_{UVW}:= Q_{UV}P_{W|U}, 
\notag \\
&\mathcal{T}_1(P_{UW}, P_{VW}):= \{P_{\tilde U \tilde V \tilde W} \in \mathcal{T}_{\Ucal \V \W} : P_{\tilde U \tilde W}=P_{UW},~ P_{\tilde V \tilde W}=P_{VW} \}, \notag 
 \end{flalign}
 \begin{flalign}
&\mathcal{T}_2(P_{UW}, P_{V}) := \{P_{\tilde U \tilde V \tilde W} \in \mathcal{T}_{\Ucal \V \W} : P_{\tilde U \tilde W}=P_{UW},~ P_{\tilde V}=P_{V}, ~ H(\tilde W| \tilde V) \geq  H_P(W|V) \}, \notag \\
 &\mathcal{T}_3(P_{UW}, P_{V}):= \{P_{\tilde U \tilde V \tilde W} \in \mathcal{T}_{\Ucal \V \W} : P_{\tilde U \tilde W}=P_{UW},~ P_{\tilde V}=P_{V} \}. \notag &&
\end{flalign}
\end{theorem}

The proof of Theorem \ref{thm:shtcceed} is given in Appendix \ref{SHTCCproof}.
Although the expression $\kappa_s(\tau)$ in Theorem \ref{thm:shtcceed} appears complicated, the terms $E_1(P_{W|U})$ to $E_4(P_{W|U},P_{SX},R,\tau)$ can be understood to correspond to distinct events that can possibly lead to a type II error. Note that
$E_1(P_{W|U})$ and $E_2(P_{W|U},P_{SX},R)$ are the same terms appearing in the error-exponent achieved by the Shimokawa et al.'s  scheme\cite{Shimokawa} for the noiseless channel setting, while $E_3(P_{W|U},P_{SX},R,\tau)$ and $E_4(P_{W|U},P_{SX},R,\tau)$ are additional terms introduced due to the noisiness of the channel. $E_3(P_{W|U},P_{SX},R,\tau)$ corresponds to the event when $M \neq 0$, $\hat M\neq M$ and $g_s^{(k)}( \hat M,V^k)=0$, whereas $E_4(P_{W|U},P_{SX},R,\tau)$ is due to the event when $M= 0$, $\hat M\neq M$ and $g_s^{(k)}( \hat M,V^k)=0$.
Note that, in general, $E_m(P_{SX})$ can take the value of $\infty$ and when this happens, the term $\tau E_m\left(P_{SX}\right)$ becomes undefined for $\tau=0$. In this case, we define $\tau  E_m\left(P_{SX}\right):=0$.
\begin{remark} \label{remanychncode}
In the SHTCC scheme, although we use Borade et al.'s  scheme for channel coding, that is concerned specifically with the protection of a special message when the ordinary message rate is $R$, any other channel coding scheme with the same rate can be employed. For instance, 
 the ordinary message can be transmitted with an error-exponent equal to the  reliability function $E(R,P_{Y|X})$ \cite{Csiszar-Korner}  of the channel $P_{Y|X}$ at rate $R$, while the special message achieves the maximum reliability possible subject to this constraint. 
However, it should be noted that a computable characterization of neither $E(R,P_{Y|X})$ (for all values of $R$) nor the associated  best reliability achievable for a single message 
is known in general. 
\end{remark}
\begin{remark} \label{remkzerratecompsch}
Similarly to the zero-rate compression scenario considered in \cite{Han} for the case of a rate-limited noiseless channel, it is possible to achieve an error-exponent of $\kappa_0(\tau)$ in general by using a one-bit communication scheme (see \cite{SD_isit19}), where 
\begin{equation}
     \kappa_0(\tau):=
    \begin{cases}
\qquad D(P_V||Q_V)   \qquad \qquad~~~,~ \mbox{if } \tau=0,\\ 
\min \left\lbrace \beta_0, \tau E_c+D(P_V||Q_V)\right\rbrace, ~  \mbox{otherwise}.
\end{cases}
\end{equation}
Here, 
\begin{align}
 \beta_0 &:=  \beta_0(P_U,P_V,Q_{UV}):=\min_{\substack{P_{\tilde U \tilde V}:\\ P_{\tilde U}=P_U,~P_{\tilde V}=P_V}}D(P_{\tilde U \tilde V}||Q_{UV}),\label{zerorateoptexp}\\
    \mbox{and } E_c&:=E_c(P_{Y|X}) := D(P_{Y|X=a}||P_{Y|X=b}), \label{chnipsachmaxdiv}
   \end{align}
   where $a$ and $b$ denote channel input symbols that satisfy
\begin{align}   
  (a,b) = \argmax_{(x,x')\in \X \times \X}  D(P_{Y|X=x}||P_{Y|X=x'}). \label{maxchannelerrexp}
     \end{align}
Note that $\beta_0$ denotes the optimal error-exponent for distributed HT over a noiseless channel, when the communication rate-constraint is zero \cite{Han}\cite{Shalaby-pap}.   
\end{remark}
In \cite{SD_isit19}, it is shown that the one-bit communication scheme mentioned in Remark \ref{remkzerratecompsch} achieves the optimal error-exponent for HT over a DMC, i.e., when the detector has no side-information. Moreover, it is also proved that optimal error-exponent is not improved if the type I error probability constraint is relaxed; and hence, strong converse holds. In the limiting case of zero channel capacity, i.e.,  $C(P_{Y|X})=0$, it is intuitive to expect that communication from the observer to the detector does not improve the achievable error-exponent for distributed HT. In Appendix \ref{proofzerocaperror-exponent} below, we show that this is indeed the case in a strong converse sense, i.e., the optimal error-exponent depends only on the side-information $V^k$, and is given by $D(P_V||Q_V)$, for any constraint $\epsilon \in (0,1)$ on the type I error probability. 
This is in contrast to the zero-rate compression case considered in \cite{Han}, where one bit of communication between the observer and detector can achieve a strictly positive error-exponent, in general.

The SHTCC  schemes introduced above  performs independent HT and channel coding, i.e., the channel encoder $f_c^{(k,n)}$ neglects $U^k$ given the output $M$ of source encoder $f_s^{(k)}$, and $g_s^{(k)}$ neglects $Y^n$ given the output of the channel decoder $g_c^{(k,n)}$. 
The following scheme ameliorates these restrictions and uses hybrid coding to  perform joint HT and channel coding.

 \subsection{JHTCC Scheme}
Hybrid coding is a form of JSCC introduced in \cite{Lim-minero-kim-2015} for the  lossy transmission of sources over noisy networks. As the name suggests, hybrid coding is a combination of the digital and analog (uncoded) transmission schemes.  For simplicity\footnote{For the case $\tau \neq 1$, as mentioned in \cite{Lim-minero-kim-2015}, we can consider hybrid coding over super symbols $U^{k^*}$ and $X^{n^*}$, where $k^*$ and $n^*$ are some integers satisfying the constraint $n^* \leq \tau k^*$. This amounts to enlarging the source and side-information r.v.'s alphabets, and thus results in a harder optimization problem over the conditional probability distributions $P_{\bar W|U^{k^*}S}$ and $P_{X^{n^*}|U^{k^*}S\bar W}$ given in Theorem \ref{thm:jhtcceed}. However, we omit its  description since the technique is standard and only adds notational clutter.}, we assume the \textit{matched-bandwidth} scenario, i.e.,  $k=n$ ($\tau=1$). In hybrid coding, the source $U^n$ is first mapped to one of the codewords $\bar{W}^n$ within a compression codebook. Then, a symbol-by-symbol function (deterministic) of the $\bar{W}^n$ and $U^n$ is transmitted as the channel codeword $X^n$. This procedure is reversed at the decoder, in which,  the decoder first attempts to obtain an estimate $\hat {\bar{W}}^n$ of $\bar{W}^n$ using the channel output $Y^n$ and its own correlated side information $V^n$. Then, the reconstruction $\hat U^n$ of the source is obtained as a symbol-by-symbol function of the reconstructed codeword, $Y^n$ and $V^n$. In this subsection, we propose a lower bound on the optimal error-exponent that is achieved by a scheme that utilizes hybrid coding for the communication between the observer and the detector, which we refer to as the JHTCC scheme. Post estimation of $\hat{\bar{W}}^n$, the detector performs the hypothesis test using $\hat{\bar{W}}^n$, $Y^n$ and $V^n$, instead of estimating $\hat U^n$ as is done in JSCC problems. We will in fact consider a slightly generalized form of hybrid coding in that the encoder and detector is allowed to perform \enquote{time-sharing} according to a sequence $S^n$ that is known a priori to both parties. Also, the input $X^n$ is allowed to be generated according to an arbitrary memoryless stochastic function instead of a deterministic function.  The JHTCC scheme will be described in detail in  Appendix \ref{JHTCCproof}. Next, we state a lower bound on $\kappa(\tau, \epsilon)$ that is achieved by the JHTCC scheme.
 \begin{theorem} \label{thm:jhtcceed}
 $\kappa(1, \epsilon) \geq \kappa_h$,~  $\forall ~\epsilon \in (0,1]$, where
\begin{align}
 \kappa_h :=  \sup_{\mathbf{b}~\in~ \mathcal{B}_h}\min \Big\{&E_1'(P_S,P_{\bar W|US},P_{X|US\bar W}),~ E_2'(P_S,P_{\bar W|US},P_{X|US\bar W}), \notag \\
&E_3'(P_S,P_{\bar W|US},P_{X'|US},P_{X|US\bar W})\Big\}, \label{hybridcodeerror-exponent} 
\end{align}
 \begin{equation*}
 \mathcal{B}_h:=\left\{
  \begin{aligned}
      &\mathbf{b}=\left(P_S,P_{\bar W|US},P_{X'|US}, P_{X|US\bar W} \right): I_{\hat P}(U;\bar W|S) <I_{\hat P}(\bar W;Y,V|S),~\X'=\X,\\
        & \hat P_{UVS\bar WX'XY}\left(P_S,P_{\bar W|US},P_{X'|US}, P_{X|US\bar W} \right):=  P_{UV}P_SP_{\bar W|US} P_{X'|US} P_{X|US\bar W} P_{Y|X}
  \end{aligned}
\right\},
\end{equation*}
\begin{align}
&E_1'\left(P_S,P_{\bar W|US},P_{X|US\bar W}\right):= \min_{P_{\tilde U \tilde V   \tilde S \tilde W \tilde Y} \in \mathcal{T}_1'\left(\hat P_{US\bar W}, \hat P_{VS\bar WY}\right) } D\left(P_{\tilde U \tilde V  \tilde S  \tilde W \tilde Y} || \hat Q_{UVS\bar WY}\right), \label{expterm1h} 
\end{align}
\begin{align}
&E_2'\left(P_S,P_{\bar W|US},P_{X|US\bar W}\right):= \min_{P_{\tilde U \tilde V \tilde S  \tilde W  \tilde Y} \in \mathcal{T}_2'\left(\hat P_{US\bar W}, \hat P_{VS\bar WY}\right) } D\left(P_{\tilde U \tilde V \tilde S  \tilde W  \tilde Y} || \hat Q_{UVS\bar WY}\right)\notag \\
& \qquad \qquad \qquad \qquad \qquad \qquad \qquad \qquad \qquad \qquad +I_{\hat P}(\bar W;V,Y|S)-I_{\hat P}(U;\bar W|S), \label{expterm2h}\\
&E_3'\left(P_S,P_{\bar W|US},P_{X'|US},P_{X|US\bar W}\right):= D(\hat P_{ V S Y} || \check Q_{VSY})+I_{\hat P}(\bar W;V,Y|S) -I_{\hat P}(U;\bar W|S), \label{expterm3h} \\
& \hat Q_{UVS\bar WX'XY}(P_S,P_{\bar W|US}, P_{X'|US},P_{X|US\bar W}):= Q_{UV}P_S P_{\bar W|US} P_{X'|US} P_{X|US\bar W} P_{Y|X},\label{jntdistalth} \\
& \check Q_{UVSX'XY}( P_{S},P_{X'|US}):= Q_{UV}P_{S}P_{X'|US} \ind (X=X') P_{Y|X},  \\
& \mathcal{T}_1'(\hat P_{US\bar W}, \hat P_{VS\bar WY}):=\{P_{\tilde U \tilde V  \tilde S  \tilde W \tilde Y} \in \mathcal{T}_{\Ucal \V \mathcal{S} \W  \Y} : P_{\tilde U \tilde S \tilde W}=\hat P_{US\bar W},~ P_{ \tilde V \tilde S  \tilde W  \tilde Y}=\hat P_{VS\bar WY} \}, \notag \\
& \mathcal{T}_2'(\hat P_{US\bar W}, \hat P_{VS\bar WY}) :=\{P_{\tilde U \tilde V \tilde S  \tilde W  \tilde Y} \in \mathcal{T}_{\Ucal \V \mathcal{S} \W   \Y} : P_{\tilde U \tilde S \tilde W}=\hat P_{US\bar W},~ P_{ \tilde V \tilde S \tilde Y}=\hat P_{VSY}, \notag \\
& \qquad \qquad  \qquad \qquad \qquad~~  H(\tilde W|\tilde V, \tilde S, \tilde Y) \geq H_{\hat P}(\bar W|V,S,Y) \}. \notag 
 \end{align}
\end{theorem}

The proof of Theorem \ref{thm:jhtcceed} is given in Appendix \ref{JHTCCproof}. 
The different factors inside the minimum in \eqref{hybridcodeerror-exponent} can be intuitively understood to be related to the various events that could possibly lead to a type 2 error. More specifically, let the event that the encoder is unsuccessful in finding a codeword $\bar{W}^n$ in the quantization codebook that is typical with $U^n$ be referred to as the \textit{encoding error}, and the event that a wrong codeword $\hat{\bar{W}}^n$ (unintended by the encoder) is reconstructed at the detector be referred to as the \textit{decoding error}. Then, $E_1'(P_S,P_{\bar W|US},P_{X|US\bar W})$ is related to the event that neither the encoding nor the  decoding error occurs, while $E_2'(P_S,P_{\bar W|US},P_{X|US\bar W})$ and $E_3'(P_S,P_{\bar W|US},P_{X'|US},P_{X|US\bar W})$ are related to the events that only the decoding error and both the encoding and decoding errors occur, respectively.
From Theorem \ref{thm:shtcceed} and Theorem \ref{thm:jhtcceed}, we have the following corollary. 
\begin{cor}
\begin{align}
    \kappa(1, \epsilon) \geq \max \left \lbrace \kappa_h, \kappa_s(1)\right\rbrace, ~ \forall \epsilon \in (0,1] .
\end{align}
\end{cor}
It is well-known that in the context of JSCC, hybrid coding recovers separate source-channel coding as a special case \cite{Lim-minero-kim-2015}. It is also known that hybrid coding, of which uncoded transmission is a special case, strictly outperforms separation based schemes in certain multi-terminal settings \cite{Cover-elgamal-salehi}. Below, we provide an example where the error-exponent achieved by the JHTCC scheme is strictly better than that achieved by the SHTCC scheme, i.e., $\kappa_h> \kappa_s(1)$. 
\begin{example} \label{jhtccopsepex}
Let $\Ucal=\V=\X=\Y=\{0,1\}$ and  $P_U=Q_U=[0.5~0.5]$. Let 
\[P_{V|U}=
\begin{bmatrix}
1-p_0 & p_0\\p_0 & 1-p_0
\end{bmatrix},
Q_{V|U}=
\begin{bmatrix}
1-p_1 &p_1\\p_1 &1-p_1
\end{bmatrix}, \mbox{ and }P_{Y|X}=
\begin{bmatrix}
1-q & q\\q & 1-q
\end{bmatrix},
\]
where $q =0.2$, $p_0=0.8$ and $p_1=0.25$. For this example, we have $ \kappa_h \geq 0.3244 >0.161 \geq \kappa_s(1)$.
\end{example}
\begin{IEEEproof}
Note that $P_V=Q_V=[0.5~ 0.5]$, and 
\begin{align}
    H_Q(V|W) \geq H_P(\bar V|W)=H_P(V |W), ~\bar V=V \oplus 1, \label{condentgrtalt}
\end{align}
for any $W$ that satisfies $V-U-W$, since\\
\[P_{\bar V|U}=
\begin{bmatrix}
1-\bar p_0 & \bar p_0\\\bar p_0 & 1-\bar p_0
\end{bmatrix}
\]
with $\bar p_0=0.2<p_1$. Then, the lower bound $\kappa_s(1)$ simplifies as
\begin{align}
    \kappa_s(1)=\sup_{\substack{(P_{W|U},P_{SX},R) \\ \in~ \mathcal{B}(1, P_{Y|X})}} \min \{ E_1(P_{W|U}),E_2(P_{W|U},P_{SX},R),E_3(P_{W|U},P_{SX},R,1)\}. \label{simpexpsepexamp}
\end{align}
To see this, consider an arbitrary $(P_{W|U},P_{SX},R) \in \mathcal{B}(1, P_{Y|X})$.
We have
\begin{flalign}
  &E_1(P_{W|U}):= \min_{P_{\tilde U \tilde V \tilde W} \in \mathcal{T}_1(P_{UW}, P_{VW}) } D(P_{\tilde U \tilde V \tilde W} ||  Q_{UVW}), \label{eqdpmarg1}\\ 
  & E_2(P_{W|U},P_{SX},R) 
 =
\begin{cases}  \label{eqdpmarg2}
R -I_P(U;W|V), \qquad \mbox{if } I_P(U;W) > R, \\  \qquad \quad   \infty, \qquad  \qquad \qquad \quad \mbox{otherwise}, 
 \end{cases} \\
 &E_3(P_{W|U},P_{SX},R,1):= 
\begin{cases} \label{eqdpmarg3}
R-I_P(U;W|V) + E_x \left(R,P_{SX}\right), \qquad \mbox{if } I_P(U;W) > R, \\
I_P(V;W) + E_x\left(R,P_{SX}\right),\qquad\qquad \qquad  \quad \mbox{otherwise}, 
\end{cases} &&
\end{flalign}
since $Q_{UVW} \in \mathcal{T}_2(P_{UW}, P_{V}) \cap \mathcal{T}_3(P_{UW}, P_{V})$, which follows from \eqref{condentgrtalt}, $P_{UW}=Q_{UW}$ and $P_V=Q_V$. This in turn implies that
\begin{align}
 \substack{ \min \\ P_{\tilde U \tilde V \tilde W} \in \mathcal{T}_2(P_{UW}, P_{V}) }~ D(P_{\tilde U \tilde V \tilde W} || Q_{UVW})=\substack{ \min \\ P_{\tilde U \tilde V \tilde W} \in \mathcal{T}_3(P_{UW}, P_{V}) }~  D(P_{\tilde U \tilde V \tilde W} || Q_{UVW})=0.   
\end{align}
Also, we have
\begin{flalign}
     E_4( P_{W|U},P_{SX},R,1) 
 &:=
\begin{cases}  \label{eqdpmarg4}
R-I_P(U;W|V)+ E_m\left(P_{SX}\right), \qquad \quad \mbox{if } I_P(U;W) > R, \\
 I_P(V;W) + E_m\left(P_{SX}\right),\qquad \qquad \qquad \qquad \mbox{otherwise,} 
 \end{cases} \\
 & \geq E_3(P_{W|U},P_{SX},R,1), &&
\end{flalign}
since $E_m(P_{SX}) \geq E_x\left(R,P_{SX}\right) $ (the reliability of a special message in Borade et al.'s scheme is at least as good as that of an ordinary message), which implies \eqref{simpexpsepexamp}. Given that \eqref{simpexpsepexamp} holds, $|\mathcal{S}|$ can be taken to be equal to 1, and $P_X$ can be chosen to be the capacity achieving channel input distribution ($P_X(0)=P_X(1)=0.5$) which maximizes $E_x\left(R,P_{SX}\right)$ (for any $R$) (see \cite{Gallager-codthm} and \cite[Exercise 10.26]{Csiszar-Korner}) without loss of generality. Hence, $I_P(X;Y)= C(P_{Y|X})=1-h_b(q)$.

Let $r:=h_b^{-1}(H_P(U|W))=h_b^{-1}(H_Q(U|W))$, where $h_b^{-1}:[0,1] \mapsto [0,0.5]$ is the inverse of the binary entropy function given by $h_b(r):=-r\log_2(r)-(1-r)\log_2(1-r)$. First, consider 
\begin{align}
P_{W|U} \in \tilde{\mathcal{B}}:=\{P_{W|U}:~I_P(U;W) < I_P(X;Y)=1-h_b(q)\}. \label{assumpsrcchnc1}    
\end{align}
Note that if $R \geq I_P(U;W)$, then $E_2(P_{W|U},P_{SX},R)=\infty$, and $E_3(P_{W|U},P_{SX},R,1)=I_P(V;W)+E_x(R,P_{SX})$. Hence, 
\begin{align}
\min \{ E_2(P_{W|U},P_{SX},R),E_3(P_{W|U},P_{SX},R,1)\} &= I_P(V;W)+E_x(R,P_{SX}) \notag \\
& \leq I_P(V;W)+E_x(I(U;W),P_{SX}), \label{exdecfnappl1}
\end{align}
where \eqref{exdecfnappl1} follows since $E_x(R,P_{SX})$ is a decreasing function of $R$. On the other hand, if $R < I_P(U;W)$, then
$E_2(P_{W|U},P_{SX},R)=R-I_P(U;W|V)$ and $E_3(P_{W|U},P_{SX},R,1) = R-I_P(U;W|V)+E_x(R,P_{SX})$
yielding that 
\begin{align}
\min \{ E_2(P_{W|U},P_{SX},R),E_3(P_{W|U},P_{SX},R,1)\} = R-I_P(U;W|V) \leq I_P(V;W). \label{compexp2twoexp}
\end{align}
Hence, from \eqref{exdecfnappl1} and \eqref{compexp2twoexp}, we have
\begin{align}
\sup_{ \substack{(P_{W|U},P_{SX},R) \in~ \mathcal{B}(1, P_{Y|X}): \\P_{W|U} \in \tilde{\mathcal{B}}}} \min \{ E_2(P_{W|U},P_{SX},R),E_3(P_{W|U},P_{SX},R,1)\} \leq I_P(V;W)+E_x(I(U;W),P_{SX}). \notag 
\end{align}
Also, note that \eqref{assumpsrcchnc1} implies  $h_b(r) \geq h_b(q)$; and hence, $r \in [q,0.5]$. Thus, we can write
\begin{align}
   I_P(V;W)+E_x(I(U;W),P_{SX}) &= 1-H_P(V|W)+ E_x(I(U;W),P_{SX}) \notag \\
   & \leq 1-h_b(h_b^{-1}(H(U|W))*p_0)+ E_x(I(U;W),P_{SX}) \label{applymrsgerblemm}\\
   &= 1-h_b(r*p_0)+ E_x(1-h_b(r),P_{SX}):=f'(r), \label{sumofconvfn}
\end{align}
where $p*q:=(1-p)q+p(1-q)$, and \eqref{applymrsgerblemm} follows by an application of Mrs. Gerber's Lemma \cite{Elgamalkim}. The plot of $f'(r)$ as a function of $r \in [q,0.5]$ is shown in Fig. \ref{explotfunc} below, which uses the expression for $E_x(R,P_{SX})$ given in \cite[Exercise 10.26]{Csiszar-Korner}. As is evident from the plot, the maximum value of  $f'(r)$ is attained at $r=0.5$, and equals $f'(0.5)=E_x(0)=-0.5*0.5*\log_2(4q(1-q))=0.161$.
It follows that
\begin{align}
\sup_{ \substack{(P_{W|U},P_{SX},R) \in~ \mathcal{B}(1, P_{Y|X}): \\P_{W|U} \in \tilde{\mathcal{B}}}} \min \{ E_2(P_{W|U},P_{SX},R),E_3(P_{W|U},P_{SX},R,1)\} \leq 0.161. \label{firstvalexexp}
\end{align}

Next, consider  that  
\begin{align}
  P_{W|U} \in \tilde {\mathcal{B}}^c:=\{P_{W|U}: I_P(W;U) \geq 1-h_b(q) \mbox{ and } I_P(U;W|V) \leq 1-h_b(q)\}. \label{secondcondex}
\end{align} 
Note that the first and  second inequalities in \eqref{secondcondex}  imply, respectively, that $r \in [0,q]$, and
\begin{align}
    1-h_b(r)-(1-h_b(r*p_0)) \leq 1-h_b(q).
\end{align}
Also, since $R< 1-h_b(q)$ holds for any $(P_{W|U},P_{SX},R) \in~ \mathcal{B}(1, P_{Y|X})$, we have $I_P(U;W) >R$, and hence,
   \begin{flalign}
   \sup_{ \substack{(P_{W|U},P_{SX},R) \in~ \mathcal{B}(1, P_{Y|X}): \\P_{W|U} \in \tilde{\mathcal{B}}^c}}
& \min\{ E_2(P_{W|U},P_{SX},R),E_3(P_{W|U},P_{SX},R,1) \}  \notag \\
 & \leq 
\sup_{ \substack{(P_{W|U},P_{SX},R) \in~ \mathcal{B}(1, P_{Y|X}): \\P_{W|U} \in \tilde{\mathcal{B}}^c}} R -I_P(U;W|V) \notag \\
 &< 1-h_b(q)-(h_b(r*p_0)-h_b(r)) \label{lasteqexmpfin1} \\
 & \leq 1-h_b(q*p_0)=0.0956, \label{lasteqexmpfin2} &&
\end{flalign} 
 \begin{figure}[t]
\centering
\includegraphics[trim=0.5cm 1cm 1.5cm 1cm, clip, width= 0.45\textwidth]{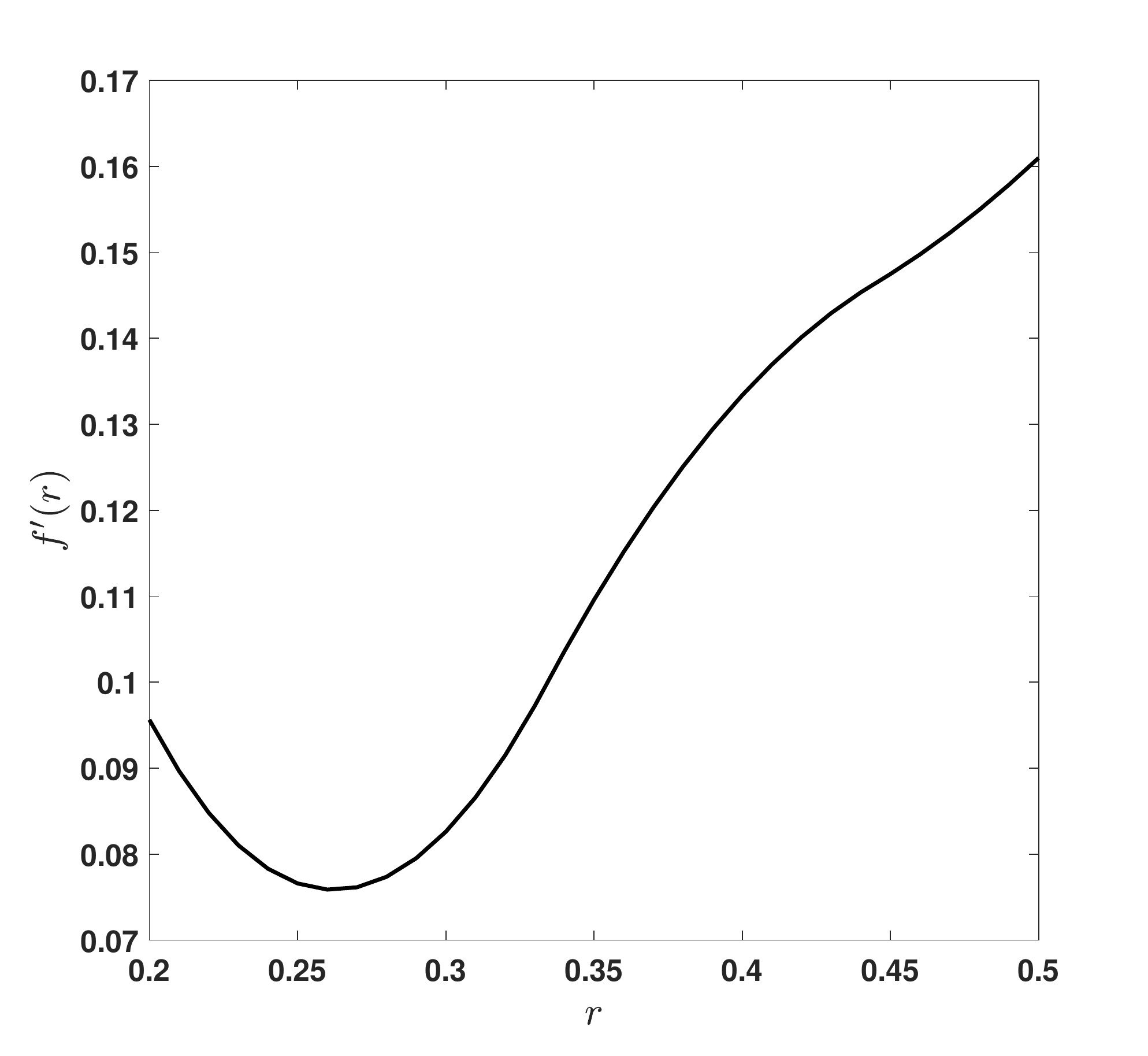}
\caption{Plot of $f'(r)$ in the range $r \in [0.2,0.5]$.} \label{explotfunc}
\end{figure}
where \eqref{lasteqexmpfin1} follows again from Mrs. Gerber's lemma, and \eqref{lasteqexmpfin2} follows since the R.H.S. of \eqref{lasteqexmpfin1} is an increasing function of $r$ and hence the maximum is attained at $r=q$ in the range $[0,q]$. Thus, from \eqref{firstvalexexp} and \eqref{lasteqexmpfin2}, it follows that $\kappa_s(1) \leq 0.161$. 

Finally, we show that the JHTCC scheme can achieve a strictly larger error-exponent, i.e.,  $\kappa_h> 0.161$. In fact,  uncoded transmission which is a special case of the JHTCC scheme with $X=X'=U$, $W=S=$ constant, achieves an error-exponent of
\begin{align}
    D(P_{VY}||Q_{VY})=D_b(q*p_0||q*p_1)=D_b(0.68||0.35)=0.3244, \label{uncodtransexp}
\end{align}
where, $D_b$ denotes the binary KL divergence defined as $D_b(p||q)= p \log_2\left(\frac{p}{q}\right)+(1-p) \log_2\left(\frac{1-p}{1-q}\right)$. Thus, we have shown that the error-exponent achieved by the JHTCC scheme is strictly greater than that achieved by the SHTCC scheme.
\end{IEEEproof}
Thus far, we obtained lower bounds on the optimal error-exponent for distributed HT over a DMC, and showed via an example that the joint scheme strictly outperforms the separation based scheme in some cases. In order to get an exact characterization of the optimal error-exponent, a matching upper bound is required. However, obtaining a tight computable upper bound  remains a challenging open problem in the general hypothesis testing case even when the channel is noiseless, and consequently, an exact computable characterization of the optimal error-exponent is unknown. However, as we show in the next section, the problem does admit single-letter characterization for TACI. 
\section{Optimality result for TACI}\label{Optresults} 
Recall that for TACI, $V=(E,Z)$ and $Q_{UEZ}=P_{UZ}P_{E|Z}$. Let
\begin{align}
\kappa(\tau)= \lim_{\epsilon \rightarrow 0} \kappa(\tau,\epsilon).  \label{error-exponentforzerot1err}   
\end{align}
We will drop the subscript $P$ from information theoretic quantities like mutual information, entropy, etc., as there is no ambiguity on the joint distribution involved, e.g., $I_{P}(U;W)$ will be denoted by $I(U;W)$. The following result holds.
\begin{prop} \label{corrsinguser}
For TACI over a DMC $P_{Y|X}$,
 \begin{equation} \label{tacierror-exponent}
 \kappa(\tau)=
\sup \left\{
  \begin{aligned}
  I(E;W|Z):&~ \exists~W \mbox{ s.t. } I(U;W|Z) \leq \tau C(P_{Y|X}),\\&
 (Z,E)-U-W, ~|\mathcal{W}| \leq |\Ucal|+1.
  \end{aligned}
\right\},~ \tau \geq 0. 
\end{equation}
\end{prop}
\begin{IEEEproof}
For the proof of achievability, we will show that $\kappa_s(\tau)$  when specialized to TACI recovers \eqref{tacierror-exponent}.   
 Let  $\mu>0$ be a arbitrarily small positive number, and
 \begin{align} 
 &\mathcal{B}'\left( \tau, P_{Y|X}\right)
:=\left\{
  \begin{aligned} \label{smallsetcond}
       (P_{W|U},P_{SX},R_m):~&\mathcal{S}=\X,~P_{UEZWSXY}(P_{W|U},P_{SX}):= P_{UEZ}P_{W|U}P_{SX}P_{Y|X},\\&~I(U;W|Z) \leq R_m:=\tau I(X;Y|S)-\mu< \tau  I(X;Y|S)  
  \end{aligned}
\right\}.
\end{align}
Note that $\mathcal{B}'(\tau, P_{Y|X}) \subseteq \mathcal{B}(\tau, P_{Y|X})$ since $I(U;W|E,Z) \leq I(U;W|Z)$, which holds due to the Markov chain $(Z,E)-U-W$. Now, consider $(P_{W|U},P_{SX},R_m) \in \mathcal{B}'(\tau,P_{Y|X})$.
Then, we have
\begin{flalign}
E_1(P_{W|U})&=  \min_{P_{\tilde U \tilde E \tilde Z \tilde W} \in \mathcal{T}_1(P_{UW}, P_{EZW}) } D(P_{\tilde  U \tilde E  \tilde Z \tilde W } || P_ZP_{U|Z}P_{E|Z}P_{W|U})   \notag \\
& \geq  \min_{P_{\tilde U \tilde E \tilde Z \tilde W} \in \mathcal{T}_1(P_{UW}, P_{EZW}) } D(P_{\tilde E  \tilde Z \tilde W } || P_ZP_{E|Z}P_{W|Z})   \label{data-processing} \\
&= I(E;W|Z), \notag  &&
\end{flalign}
where \eqref{data-processing} follows from the log-sum inequality \cite{Csiszar-Korner}. Also,
\begin{flalign}
& E_2\left(P_{W|U},P_{SX},R_m\right)\geq R_m-I(U;W|E,Z) \geq  I(U;W|Z)-I(U;W|E,Z) = I(E;W|Z), \notag \\[5 pt]
& \substack{ \min \\ P_{\tilde U \tilde E \tilde Z \tilde W} \in \mathcal{T}_3(P_{UW}, P_{EZ}) }~  D(P_{\tilde U \tilde E \tilde Z \tilde W} || P_ZP_{U|Z}P_{E|Z}P_{W|U}) + R_m-I(U;W|E,Z)  +\tau E_x\left( \frac{R_m}{\tau},P_{SX}\right) \notag  \\
& \geq   I(U;W|Z)-I(U;W|E,Z)= I(E;W|Z), \label{recovertaci1} &&
\end{flalign}
\begin{flalign}
&\substack{ \min \\ P_{\tilde U \tilde E \tilde Z \tilde W} \in \mathcal{T}_3(P_{UW}, P_{EZ}) }~  D(P_{\tilde U \tilde E \tilde Z \tilde W} ||  P_ZP_{U|Z}P_{E|Z}P_{W|U})+I(E,Z;W) +\tau E_x\left(\frac{R_m}{\tau},P_{SX}\right)  \notag \\
&\geq  I(E;W|Z), \label{recovertaci2} \\[5 pt]
& D(P_{ EZ} || P_{EZ}) +R_m-I(U;W|E,Z) +\tau E_m\left(P_{SX}\right)   \geq  I(U;W|Z)-I(U;W|E,Z)= I(E;W|Z), \label{recovertaci3} \\
& D(P_{ EZ} || P_{EZ}) +I(E,Z;W) +\tau E_m\left(P_{SX}\right) \geq I(E;W|Z), \label{recovertaci4} &&
\end{flalign}
where in \eqref{recovertaci1}-\eqref{recovertaci4}, we used the non-negativity of KL-divergence, $E_x(\cdot, \cdot)$ and $E_m(\cdot)$. Thus, from \eqref{recovertaci1}-\eqref{recovertaci4}, it follows that
\begin{align}
    &E_3(P_{W|U},P_{SX},R_m,\tau) \geq I(E;W|Z), \label{thirdexplargtaci} \\
 \mbox{and }   &E_4(P_{W|U},P_{SX},R_m, \tau) \geq I(E;W|Z). \label{fourthexplargtaci}
\end{align}
Denoting $\mathcal{B}(\tau,P_{Y|X})$ and $\mathcal{B}'(\tau,P_{Y|X})$ by $\mathcal{B}$ and $\mathcal{B}'$, respectively, we obtain
\begin{flalign}
&\kappa(\tau, \epsilon) \notag \\
&\geq  \sup_{(P_{W|U},P_{SX},R_m) \in \mathcal{B}}  \min \Big(E_1(P_{W|U}), E_2(P_{W|U},P_{SX},R_m),E_3(P_{W|U},P_{SX},R_m,\tau), E_4(P_{W|U},P_{SX},R_m,\tau)\Big) \notag \\
& \geq  \sup_{(P_{W|U},P_{SX},R_m) \in \mathcal{B}} I(E;W|Z) \notag \\
& \geq  \sup_{(P_{W|U},P_{SX},R_m) \in \mathcal{B}'} I(E;W|Z) \label{maxsmset}\\
& =  \sup_{P_{W|U}: I(W;U|Z) \leq \tau C(P_{Y|X})-\mu}  I(E;W|Z), \label{eqclosure}&&
\end{flalign}
where \eqref{maxsmset} follows from the fact that $\mathcal{B}' \subseteq \mathcal{B}$; and \eqref{eqclosure} follows by maximizing over all $P_{SX}$ and noting that $\underset{P_{XS}}{\sup}~ I(X;Y|S)=C(P_{Y|X})$. The proof of achievability is complete by noting that $\mu>0$ is arbitrary and $I(E;W|Z)$ and $I(U;W|Z)$ are continuous functions of $P_{W|U}$. \\
\textit{Converse:}
For any sequence of encoding functions $f^{(k,n_k)}$, acceptance regions $\mathcal{A}_{(k,n_k)}$ for $H_0$ such that $n_k \leq \tau k$ and
\begin{align}
    \limsup_{k \rightarrow \infty} \alpha\left(k,n_k, f^{(k,n_k)}, g^{(k,n_k)} \right) =0, \label{vanisht1prconst}
\end{align} 
we have similar to \cite[Theorem 1 (b)]{Ahlswede-Csiszar}, that
\begin{flalign}
\limsup_{k \rightarrow \infty}  \frac{-1}{k}\log \left( \beta \left(k,n_k, f^{(k,n_k)}, g^{(k,n_k)} \right) \right) &  \leq \limsup_{k \rightarrow \infty} \frac 1k D\left(P_{Y^{n_k}E^kZ^k}|| Q_{Y^{n_k}E^kZ^k}\right) \label{divaserror-exponent} \\
& = \limsup_{n \rightarrow \infty} ~ \frac 1k I(Y^{n_k};E^k|Z^k) \label{kldivtomutinf}\\
&=H(E|Z)-\liminf_{k \rightarrow \infty} \frac 1k H(E^k|Y^{n_k},Z^k), \label{finalbndexpsing} &&
\end{flalign}
where \eqref{kldivtomutinf} follows since $Q_{Y^{n_k}E^kZ^k}= P_{Y^{n_k}Z^k}P_{E^k|Z^k}$. Now, let $T$ be a r.v. uniformly distributed over $[k]$ and independent of all the other r.v.'s $(U^k,E^k,Z^k,X^{n_k},Y^{n_k})$. Define an auxiliary r.v. $W := (W_{T},T)$, where $W_{i} := (Y^{n_k},E^{i-1},Z^{i-1},Z_{i+1}^k)$, $i \in [k]$. Then, the last term can be single-letterized as follows.
\begin{flalign}
 H(E^k|Y^{n_k},Z^k) &=\sum\nolimits_{i=1}^k H(E_i|E^{i-1},Y^{n_k},Z^k) \notag \\
 &=\sum\nolimits_{i=1}^k  H(E_i|Z_i,W_{i})  \notag \\
 &= k H(E_T|Z_T,W_{T},T) \notag \\
 &=k H(E|Z,W). \label{rtbndsrc}&&
\end{flalign}
Substituting \eqref{rtbndsrc} in \eqref{finalbndexpsing}, we obtain
\begin{align}
   \limsup_{k \rightarrow \infty}  \frac{-1}{k}\log \left( \beta \left(k,n_k, f_{1}^{(k,n_k)}, g^{(k,n_k)} \right) \right) \leq I(E;W|Z).
\end{align}
Next, note that the data processing inequality applied to the Markov chain $(Z^k,E^k)-U^k-X^n-Y^n$ yields $I(U^k;Y^{n_k}) \leq I(X^{n_k};Y^{n_k})$ which implies that
\begin{align}
I(U^k;Y^{n_k})-I(U^k;Z^k) \leq  I(X^{n_k};Y^{n_k}). \label{chanupbnd}
\end{align}
The R.H.S. of \eqref{chanupbnd} can be upper bounded due to the memoryless nature of the channel as
\begin{align}
    I(X^{n_k};Y^{n_k}) \leq n_k \max_{P_X}I(X;Y)=n_kC(P_{Y|X}),
\end{align}
while the left hand side (L.H.S.) can be simplified as follows.
\begin{flalign}
  I(U^k;Y^{n_k})-I(U^k;Z^k)&= I(U^k;Y^{n_k}|Z^k)  \label{MP1}\\
  &=\sum\nolimits_{i=1}^k I(Y^{n_k};U_{i}|U^{i-1},Z^k) \notag \\
  &= \sum\nolimits_{i=1}^k I(Y^{n_k},U^{i-1},Z^{i-1},Z_{i+1}^k;U_i|Z_i) \label{memorylessprop} \\
    &= \sum\nolimits_{i=1}^k I(Y^{n_k},U^{i-1},Z^{i-1},Z_{i+1}^k,E^{i-1};U_i|Z_i) \label{markprophyp} \\
  & \geq  \sum\nolimits_{i=1}^k I(Y^{n_k},Z^{i-1},Z_{i+1}^k,E^{i-1};U_i|Z_i) \notag  \\
& =  \sum\nolimits_{i=1}^k I(W_{i};U_i|Z_i) = kI(W_T;U_T|Z_T,T) \notag   \\
&=k I(W_T,T;U_T|Z_T) \label{auxtimeshar} \\
&= k I(W;U|Z). \notag &&
\end{flalign}
Here, \eqref{MP1} follows due to $Z^k-U^k-Y^{n_k}$; \eqref{memorylessprop} follows since the sequences $(U^k,Z^k)$ are memoryless; 
  \eqref{markprophyp} follows since $E^{i-1}-(Y^{n_k},U^{i-1},Z^{i-1},Z_{i+1}^k)-U_i$ ; \eqref{auxtimeshar} follows from the fact that $T$ is independent of all the other r.v.'s.
Finally, note that $(E,Z)-U-W$ holds and that the cardinality bound on $W$ follows by standard arguments based on Caratheodory's theorem. This completes the proof of the converse, and hence of the proposition. 
\end{IEEEproof}

As the above result shows, TACI is an instance of distributed HT over a DMC, in which, the optimal error-exponent is equal to that achieved over a noiseless channel of the same capacity. Hence, a noisy channel does not always degrade the achievable error-exponent. Also, notice that a separation based coding scheme that performs independent HT and channel coding is sufficient to achieve the optimal error-exponent for TACI. The investigation of a single-letter characterization of the optimal error-exponent for TACI over a DMC is inspired from an analogous result for TACI over a noiseless channel. It would be interesting to explore whether the noisiness of the channel enables obtaining computable characterizations of the error-exponent for some other special cases of the problem.
  \section{Concluding remarks}
 \label{sec:conclu}
In this paper, we have studied the error-exponent achievable for  distributed HT problem over a DMC with side information available at the detector. We obtained single-letter lower bounds on the optimal error-exponent for general HT, and exact single-letter characterization for TACI. It is interesting to note from our results that the reliability function of the channel does not play a role in the characterization of the optimal error-exponent for TACI, and only the channel capacity matters. We also showed via an example that the lower bound on the error-exponent obtained using our joint hypothesis testing and channel coding scheme is strictly better than that obtained using our separation based scheme. Although this does not imply that \enquote{separation does not hold} for distributed HT over a DMC, it points to the possibility that joint HT and channel coding schemes outperform separation based schemes, in general, and it is worthwhile investigating this aspect in greater detail. While a strong converse holds for distributed HT over a rate-limited noiseless channel \cite{Ahlswede-Csiszar}, it remains an open question whether this property holds for noisy channels. As a first step, it is shown in \cite{SD_isit19} that this is indeed the case for HT over a DMC with no side-information. While we did not discuss the complexity of the schemes considered in this paper, it is an important factor that needs to be taken into account in any practical implementation of these schemes. In this regard, it is evident that the SHTCC and JHTCC schemes are in increasing order of complexity.
\begin{appendices} 
\section{Proof of Theorem \ref{thm:shtcceed}} \label{SHTCCproof}
The proof outline is as follows. We first describe the encoding and decoding operations of the SHTCC scheme. The random coding method is used to analyze the type I and type II error probabilities achieved by this scheme, averaged over the ensemble of randomly generated codebooks. By the standard expurgation  technique \cite{Gallager-codthm} (e.g.,  removing \enquote{worst} codebooks in the ensemble  with the highest type I error probability such that the total probability of the removed codebooks lies in the interval $ (0.5,1)$), this guarantees the existence of at least one deterministic codebook that achieves type I and type II error probabilities of the same order, i.e., within a constant multiplicative factor. Since, in our scheme below, the type I error probability averaged over the random code ensemble vanishes asymptotically with the the number of samples $k$, the same holds for the  codebook obtained after expurgation. Moreover, the error-exponent is not affected by a constant multiplicative factor on the type II error probability, and thus, this codebook asymptotically achieves  the same type I error probability and error-exponent as the average.

For brevity, in the proof below, we denote the information theoretic quantities like $I_{P}(U;W)$, $T_{[P_{UW}]_{\delta}}^k$, etc., that are computed with respect to joint distribution $P_{UVWSXY}$ given in \eqref{probdistsepact} below by  $I(U;W)$, $T_{[UW]_{\delta}}^k$, etc.

\textit{Codebook Generation}: Let $k \in \mathbb{Z}^+$ and $n=\lfloor \tau k \rfloor$. Fix  a finite alphabet $\mathcal{W}$, a positive number (small) $\delta>0$, and distributions $P_{W|U}$ and $P_{SX}$. Let $\delta':=\frac{\delta}{2},~\hat\delta := |\Ucal|\delta, ~\tilde \delta:=2 \delta,~ \bar{\delta}:= \frac{\delta'}{|\V|},~\check{\delta}:=|\W|\tilde \delta$ and
\begin{align}
    P_{UVWSXY}(P_{W|U},P_{SX}):= P_{UV}P_{W|U}P_{SX}P_{Y|X}. \label{probdistsepact}
\end{align}
 Let $ \mu=O(\delta)$ (subject to  constraints that will be specified below) and  $R$ be such that 
\begin{align}
   I(U;W|V)+2\mu  \leq R \leq \tau I(X;Y|S)- \mu. \label{rateconstchosen} 
\end{align}
Denoting $M_k':= e^{k(I(U:W)+\mu)}$, the \textit{source codebook} $\mathcal{C}$ used by the source encoder $f_s^{(k)}$ is obtained by generating $M_k'$ sequences $w^k(j), ~j \in \left[M_k'\right]$, independently at random according to the distribution $\prod_{i=1}^k P_W(w_i)$, where  
\begin{align}
P_{W}(w)= \sum_{u \in \Ucal} P_{W|U}(w|u)P_U(u), \forall~w \in \W. \notag
\end{align}
The \textit{channel codebook} $\tilde {\mathcal{C}}$ used by $f_c^{(k,n)}$ is obtained as follows. The codeword length $n$ is divided into $|\mathcal{S}|=|\mathcal{X}|$ blocks, where the length of the first block is $\lceil P_S(s_1)n \rceil$, the second block is $\lceil P_S(s_2)n \rceil$, so on so forth, and the length of the last block is chosen such that the total length is $n$. The codeword $x^n(0)=s^n$ corresponding to $M=0$ is obtained by repeating the letter $s_i$ in block $i$. The remaining $\left \lceil e^{kR} \right \rceil $  ordinary codewords $x^n(m), ~m \in \left[  e^{kR} \right]$, are obtained by blockwise i.i.d. random coding, i.e., the symbols in the $i^{th}$ block of each codeword are generated i.i.d. according to $ P_{X|S=s_i}$. The sequence $s^n$ is revealed to the detector. 

\textit{Encoding}:
If $I(U;W)+\mu > R$, i.e., the number of codewords in the source codebook is larger than the number of codewords in the channel codebook, the encoder performs uniform random binning on the sequences $w^k(i), ~i \in \left[M_k' \right]$ in $\mathcal{C}$, i.e.,  for each codeword in $\mathcal{C}$, it selects an index uniformly at random from the set  $[e^{kR}]$. Denote the bin index selected for $w^k(i)$ by $f_B(i)$.
If the observed sequence $U^k=u^k$ is typical, i.e., $u^k \in T_{[U]_{\delta'}}^k$, the source encoder $f_s^{(k)}$ first looks for a sequence $w^k(j)$ in $\mathcal{C}$ such that $(u^k, w^k(j)) \in T_{[UW]_{\delta}}^k$. If there exist multiple such codewords, it chooses an index $j$ among them uniformly at  random, and outputs the bin-index $ M=m=f_B(j)$, $m \in [e^{kR}]$ or $M=m=j$ depending on whether $I(U;W)+\mu > R$, or otherwise.
If $u^k \notin T_{[U]_{\delta'}}^k$ or such an index $j$ does not exist, $f_s^{(k)}$ outputs the \textit{error} message $M=0$. The channel encoder $f_c^{(k,n)}$ transmits the codeword $x^n(m)$ from codebook $\tilde{\mathcal{C}}$. 

\textit{Decoding}: At the decoder, $g_c^{(k,n)}$ outputs $\hat M=0$ if for some $1 \leq i \leq |\mathcal{S}|$,
 the channel outputs corresponding to the $i^{th}$ block does not belong to  $T_{[P_{Y|S=s_i}]_{\delta}}^n$.  Otherwise, $\hat M$ is set as the index of the codeword  corresponding to the maximum-likelihood candidate among the ordinary codewords. 
If $\hat M=0$, $H_1$ is declared. Else, given the side information sequence $V^k=v^k$ and estimated bin-index $\hat M=\hat m$, $g_s^{(k,n)}$ searches for a typical sequence $ \hat w^k=w^k(\hat j) \in T_{[W]_{\hat\delta}}^k$,  in codebook $\mathcal{C}$ such that
\begin{align}
 \hat j&= \argmin_{ \substack{l:~f_B(l)=\hat m, \\  w^k(l)\in T_{[W]_{\hat\delta}}^k}} H_e(w^k(l)|v^k), \mbox{ if }I(U;W)+\mu > R ,\notag\\
  \hat j&=\hat m,\mbox{ otherwise}. \notag
\end{align}
 The decoder declares $\hat H=0$ if $(\hat w^k,v^k) \in T_{[WV]_{\tilde \delta}}^k$. Else, $\hat H=1$ is declared. 

We next analyze the type I and type II error probabilities achieved by the above scheme. 

\textbf{Analysis of Type I error}: A type I error occurs only if one of the following events happen.
\begin{flalign}
\mathcal{E}_{TE}&= \left\lbrace (U^k,V^k) \notin T_{[UV]_{\bar \delta}}^k\right\rbrace \notag\\
\mathcal{E}_{EE}&=  \left\lbrace \nexists ~j \in  \left[M_k'\right]:  (U^k,W^k(j)) \in T_{[UW]_{\delta}}^k\right\rbrace \notag \\
\mathcal{E}_{ME}&= \left\lbrace (V^k,W^k(J)) \notin T_{[VW]_{\tilde \delta}}^k \right\rbrace \notag \\
\mathcal{E}_{DE}&=   \Bigg\{ \exists ~ l \in  \left[M_k'\right],~l \neq J: f_B(l)=f_B(J), ~W^k(l) \in T_{[W]_{\hat\delta}}^k, ~H_e(W^k(l)|V^k) \leq H_e(W^k(J)|V^k) \Bigg\} \notag \\
\mathcal{E}_{CD}&= \left\lbrace g_c^{(k,n)}(Y^n) \neq M \right\rbrace \notag &&
\end{flalign}
 $\mathbb{P}(\mathcal{E}_{TE}|H=0)$ tends to $0$ asymptotically by the weak law of large numbers. Conditioned on $\mathcal{E}_{TE}^c$, $U^k \in T_{[U]_{\delta'}}$ and by the covering lemma \cite[Lemma 9.1]{Csiszar-Korner}, it is well known that for $\mu=O(\delta)$ chosen appropriately, $\mathbb{P}(\mathcal{E}_{EE}|\mathcal{E}_{TE}^c)$ tends to $0$ doubly exponentially with $k$. Given $\mathcal{E}_{EE}^c \cap \mathcal{E}_{TE}^c$ holds, it follows from the Markov chain relation $V-U-W$ and the Markov lemma \cite{Elgamalkim}, that $\mathbb{P}(\mathcal{E}_{ME}|\mathcal{E}_{TE}^c \cap \mathcal{E}_{EE}^c)$ tends to zero as $k \rightarrow \infty$. 
 Next, we consider  $\mathbb{P}(\mathcal{E}_{DE})$. 
Given that $\mathcal{E}_{ME}^c \cap  \mathcal{E}_{EE}^c \cap \mathcal{E}_{TE}^c$ holds, note that for $k$ sufficiently large, $ H_e(W^k(J)|V^k) \leq  H(W|V)+O(\delta)$. Thus, we have (for sufficiently large $k$) 
\begin{flalign}
& \mathbb{P}(\mathcal{E}_{DE}|  ~~V^k=v^k,W^k(J)=w^k,\mathcal{E}_{ME}^c \cap \mathcal{E}_{EE}^c \cap \mathcal{E}_{TE}^c) \notag \\
&\leq \sum_{\substack{l=1, \\ l \neq J}}^{M_k'} ~\sum_{ \substack{\tilde w^k \in T_{[W]_{\hat \delta}}^k:\\H_e(\tilde w^k|v^k) \\ \leq H_e(w^k|v^k)}}~ \mathbb{P}\Big( f_B(l)=f_B(J), ~~W^k(l) =\tilde w^k| ~V^k=v^k,W^k(J)=w^k,  \mathcal{E}_{ME}^c \cap \mathcal{E}_{EE}^c \cap \mathcal{E}_{TE}^c \Big) \notag \\
& = \sum_{\substack{l=1, \\ l \neq J}}^{M_k'}  \sum_{ \substack{\tilde w^k \in T_{[W]_{\hat \delta}}^k :\\H_e(\tilde w^k|v^k) \leq H_e(w^k|v^k)}} \mathbb{P}(W^k(l)=\tilde w^k|~V^k=v^k,W^k(J)=w^k,\mathcal{E}_{ME}^c \cap \mathcal{E}_{EE}^c \cap \mathcal{E}_{TE}^c)~ \frac{1}{e^{kR}}\notag\\
& \leq \sum_{\substack{l=1, \\ l \neq J}}^{M_k'} \sum_{ \substack{\tilde w^k \in T_{[W]_{\hat \delta}}^k: \\H_e(\tilde w^k|v^k) \leq H_e(w^k|v^k)}} 2 \cdot e^{-kR} e^{-k(H(W)-O(\delta))}\label{eqtwicebndprob} &&
\end{flalign}
\begin{flalign}
& \leq \sum_{\substack{l=1, \\ l \neq J}}^{M_k'} (k+1)^{|\mathcal{V}||\mathcal{W}|}~ e^{k(H(W|V)+O(\delta))} \cdot  2 \cdot e^{-kR} e^{-k(H(W)-O(\delta))} \label{simpeq1} \\
& \leq e^{-k(R-I(U;W|V)-\delta_1^{(k)})}, \label{type1errfinexpbin} &&
\end{flalign}
where 
\begin{align}
   \delta_1^{(k)} = \mu+O(\delta)+\frac{1}{k}|\V||\W|\log(k+1)+ \frac{\log(2)}{k}. \notag 
\end{align}
 To obtain \eqref{eqtwicebndprob}, we used the fact that
\begin{align}
& \mathbb{P}(W^k(l)=\tilde w^k|~\mathcal{E}_{ME}^c \cap \mathcal{E}_{EE}^c \cap \mathcal{E}_{TE}^c, W^k(J)=w^k, V^k=v^k) \leq 2 \cdot \mathbb{P}(W^k(l)=\tilde w^k).
\end{align}
This follows similarly to \eqref{bndtwiceprob}, which is discussed in the type II error analysis section below. In order to obtain the expression in \eqref{simpeq1}, we first summed over the types $ P_{\tilde W}$ of sequences within the typical set $T_{[W]_{\delta}}^k$ that have empirical entropy less than $H_e(w^k|v^k)$; and used the facts that the number of sequences within such a type is upper bounded by $e^{k(H(W|V)+\gamma_1(k))}$, 
and the total number of types is upper bounded by $(k+1)^{|\mathcal{V}||\mathcal{W}|}$ \cite{Csiszar-Korner}. Summing over all $(w^k,v^k) \in T_{[VW]_{\tilde \delta}}^k$, we obtain (for sufficiently large $k$) that
\begin{flalign}
& \mathbb{P}(\mathcal{E}_{DE}|\mathcal{E}_{ME}^c \cap \mathcal{E}_{EE}^c \cap \mathcal{E}_{TE}^c) \notag \\
&\leq \sum_{(w^k,v^k) \in T_{[WV]_{\tilde \delta}}^k} \mathbb{P}(W^k(J)=w^k,V^k=v^k|\mathcal{E}_{ME}^c \cap \mathcal{E}_{EE}^c \cap \mathcal{E}_{TE}^c)~ e^{-k(R-I(U;W|V)-\delta_1^{(k)})} \notag \\
&\leq e^{-k(R-I(U;W|V)-\delta_1^{(k)})} \leq e^{-k\frac{\mu}{2}}, \label{expdectype1err} &&
\end{flalign}
where, \eqref{expdectype1err} follows from \eqref{rateconstchosen} by choosing $\mu=O(\delta)$ appropriately.

Finally, we consider the event $\mathcal{E}_{CD}$. 
Denoting by $\mathcal{E}_{CT}$, the event that the channel outputs corresponding to the  $i^{th}$ block does not belong to $T_{[P_{Y|S=s_i}]_{\delta}}^n$ for some $1 \leq i \leq |\mathcal{S}|$, it follows from the weak law of large numbers and the union bound,  that 
\begin{align}
    \mathbb{P}(\mathcal{E}_{CT}|\mathcal{E}_{EE}^c) \xrightarrow{(k)} 0.
\end{align}
 Also, it follows from \cite[Exercise 10.18, 10.24]{Csiszar-Korner} that for sufficiently large $n$ (depending on $\mu$, $\tau,~|\X|$ and $|\Y|$),
\begin{align}
    \mathbb{P} \left( \mathcal{E}_{CD}| \mathcal{E}_{EE}^c \cap \mathcal{E}_{CT}^c\right) \leq e^{-n E_x(\frac{R}{\tau}+\frac{\mu}{2 \tau},P_{SX})}. \label{chndecerrexp}
\end{align}
 This implies that the probability 
that an error occurs at the channel decoder $g_c^{(k,n)}$ tends to $0$ as $n \rightarrow \infty$ since $E_x(\frac{R}{\tau}+\frac{\mu}{2 \tau},P_{SX})>0$ for $R \leq \tau I(X;Y|S)-\mu$.
Thus, since $ I(U;W|V)+\mu \leq  R \leq \tau I(X;Y|S)-\mu$, the probability of the events causing  type I error tends to zero asymptotically.

\textbf{Analysis of Type II error}: First, note that a type II error occurs only if $V^k \in T_{[V]_{\check{\delta}}}^k$, and hence, we can restrict the type II error analysis to only such $V^k$. Denote the event that a type II error happens by $\mathcal{D}_0$. Let
\begin{align}
\mathcal{E}_{0}= \left\lbrace U^k \notin T_{[U]_{\delta'}}^k \right\rbrace. \label{encerrevnt}
\end{align}
Then, the type II error probability can be written as
\begin{flalign}
&\beta\left(k,n, f^{(k,n)},g^{(k,n)}\right) \notag \\
&= \sum_{(u^k,v^k) \in \Ucal^k \times \V^k} \mathbb{P}(U^k=u^k,V^k=v^k|H=1) ~\mathbb{P}(\mathcal{D}_0|U^k=u^k,V^k=v^k). \label{firsteqspl}&&
\end{flalign}
Let $\mathcal{E}_{NE} :=  \mathcal{E}_{EE}^c \cap \mathcal{E}_{0}^c$. The last term in \eqref{firsteqspl} can be upper bounded as follows.

 \begin{flalign}
&\mathbb{P}(\mathcal{D}_0|U^k=u^k,V^k=v^k) \notag \\
&= \mathbb{P}(\mathcal{E}_{NE}|U^k=u^k,V^k=v^k) ~\mathbb{P}(\mathcal{D}_0|U^k=u^k,V^k=v^k,\mathcal{E}_{NE}) \notag \\
&\qquad + \mathbb{P}(\mathcal{E}_{NE}^c |U^k=u^k,V^k=v^k) ~\mathbb{P}(\mathcal{D}_0|U^k=u^k,V^k=v^k,\mathcal{E}_{NE}^c) \notag \\
& \leq \mathbb{P}(\mathcal{D}_0|U^k=u^k,V^k=v^k,\mathcal{E}_{NE})+\mathbb{P}(\mathcal{D}_0|U^k=u^k,V^k=v^k,\mathcal{E}_{NE}^c). \notag &&
\end{flalign}
Thus, we have
\begin{flalign}
&\beta\left(k,n, f^{(k,n)},g^{(k,n)}\right)\notag \\
&\leq \sum_{\substack{(u^k,v^k) \\\in~\Ucal^k \times \V^k}} \mathbb{P}(U^k=u^k,V^k=v^k|H=1)\Big[\mathbb{P}(\mathcal{D}_0|U^k=u^k,V^k=v^k,\mathcal{E}_{NE})\notag \\
&  \qquad \qquad \qquad \qquad \qquad \qquad \qquad \qquad \qquad +\mathbb{P}(\mathcal{D}_0|U^k=u^k,V^k=v^k,\mathcal{E}_{NE}^c)\Big]. \label{finsplitfirst}&&
\end{flalign}
First, we assume that $\mathcal{E}_{NE}$ holds. Then,
\begin{flalign}
\mathbb{P}(\mathcal{D}_0|~ U^k=u^k,V^k=v^k,\mathcal{E}_{NE})&=  \sum_{j=1}^{M_k'} \sum_{m=1}^{e^{kR}}\mathbb{P}(J=j, f_B(J)=m |~U^k=u^k,V^k=v^k,\mathcal{E}_{NE}) \notag \\
&\quad \quad  \qquad \qquad  \mathbb{P}(\mathcal{D}_0|U^k=u^k,V^k=v^k,J=j, f_B(J)=m,~\mathcal{E}_{NE}). \label{t2maineq}&&
\end{flalign}

By the symmetry of the codebook generation, encoding and decoding procedure, the term  
 $ \mathbb{P}(\mathcal{D}_0|U^k=u^k,V^k=v^k,J=j, f_B(J)=m, ~\mathcal{E}_{NE})$ in \eqref{t2maineq} is independent of the value of $J$ and $f_B(J)$. Hence, w.l.o.g. assuming $J=1$ and $f_B(J)=1$, we can write
\begin{flalign}
&\mathbb{P}(\mathcal{D}_0|~U^k=u^k,V^k=v^k,\mathcal{E}_{NE}) \notag \\
&= \sum_{j=1}^{M_k'} \sum_{m=1}^{e^{kR}} \mathbb{P}(J=j, f_B(J)=m |~U^k=u^k,V^k=v^k,\mathcal{E}_{NE})   \mathbb{P}(\mathcal{D}_0|U^k=u^k,V^k=v^k,J=1, f_B(J)=1,~\mathcal{E}_{NE})\notag \\
&=\mathbb{P}(\mathcal{D}_0|U^k=u^k,V^k=v^k,J=1,f_B(J)=1,~\mathcal{E}_{NE})\notag \\
&=\sum_{\substack{w^k \in \W^k }} \mathbb{P}(W^k(1)=w^k |U^k=u^k,V^k=v^k,J=1,f_B(J)=1, ~\mathcal{E}_{NE}) \notag \\
& \qquad \qquad \qquad \quad \mathbb{P}(\mathcal{D}_0|U^k=u^k,V^k=v^k,J=1,f_B(J)=1,W^k(1)=w^k,~\mathcal{E}_{NE}). \label{lsttermexp}&&
\end{flalign}

Given $\mathcal{E}_{NE}$ holds, $\mathcal{D}_0$ may occur in three possible ways: (i)  when $\hat M \neq 0$, i.e., $\mathcal{E}_{CT}^c$ occurs, the channel decoder makes an error and the codeword retrieved from the bin is jointly typical with $V^k$; (ii) when an unintended wrong codeword is retrieved from the correct bin that is jointly typical with $V^k$; and (iii) when there is no error at the channel decoder and the correct codeword is retrieved from the bin, that is also jointly typical with $V^k$.  We refer to the event in case (i) as the \textit{channel error event} $\mathcal{E}_{CE}$, and the one in case (ii) as the \textit{binning error event}  $\mathcal{E}_{BE}$. More specifically,
\begin{flalign}
\mathcal{E}_{CE}&= \{ \mathcal{E}_{CT}^c \mbox{ and }\hat M = g_c^{(k,n)}(Y^n) \neq M \},\label{chnerrevnt} \\
\mbox{and }\mathcal{E}_{BE}&=\Big\{ \exists ~ l \in  \left[M_k'\right],~ l\neq J,~ f_B(l)= \hat M,~W^k(l)) \in T_{[W]_{\hat  \delta}}^k, (V^k,W^k(l)) \in T_{[VW]_{\tilde\delta}}^k \Big\}. \label{binerrevnt} &&
\end{flalign}
Define the following events
\begin{flalign}
\mathcal{F}&= \{U^k=u^k,V^k=v^k,J=1,f_B(J)=1,W^k(1)=w^k,~\mathcal{E}_{NE}\}, \label{ev1} \\
\mathcal{F}_1&= \{U^k=u^k,V^k=v^k,J=1,f_B(J)=1,W^k(1)=w^k,~\mathcal{E}_{NE},~\mathcal{E}_{CE}\}, \label{ev2} \\
\mathcal{F}_2&= \{U^k=u^k,V^k=v^k,J=1,f_B(J)=1,W^k(1)=w^k,~\mathcal{E}_{NE},~ \mathcal{E}_{CE}^c\}, \label{ev3} \\
\mathcal{F}_{21}&= \{U^k=u^k,V^k=v^k,J=1,f_B(J)=1,W^k(1)=w^k,~\mathcal{E}_{NE}, ~\mathcal{E}_{CE}^c,~\mathcal{E}_{BE}\}, \label{ev4} \\
\mathcal{F}_{22}&= \{U^k=u^k,V^k=v^k,J=1,f_B(J)=1,W^k(1)=w^k,~\mathcal{E}_{NE}, ~\mathcal{E}_{CE}^c,~\mathcal{E}_{BE}^c\}. \label{ev5}&&
\end{flalign}
The last term in \eqref{lsttermexp} can be expressed as follows.
\begin{align}
 &\mathbb{P}(\mathcal{D}_0|\mathcal{F}) = \mathbb{P}(\mathcal{E}_{CE}|\mathcal{F}) ~\mathbb{P}(\mathcal{D}_0|\mathcal{F}_1) + \mathbb{P}(\mathcal{E}_{CE}^c|\mathcal{F})~ \mathbb{P}(\mathcal{D}_0|\mathcal{F}_2), \notag
\end{align}
where
\begin{align}
 &\mathbb{P}(\mathcal{D}_0|\mathcal{F}_2) =  \mathbb{P}(\mathcal{E}_{BE}|\mathcal{F}_2) ~ \mathbb{P}(\mathcal{D}_0|\mathcal{F}_{21}) + \mathbb{P}(\mathcal{E}_{BE}^c|\mathcal{F}_2) ~ \mathbb{P}(\mathcal{D}_0|\mathcal{F}_{22}). \label{maineqspl2}
\end{align}
It follows from \eqref{chndecerrexp} that for sufficiently large $k$,
\begin{align}
 \mathbb{P}(\mathcal{E}_{CE}|\mathcal{F})& \leq e^{-n E_x(\frac{R}{\tau}+\frac{\mu}{2 \tau},P_{SX})} 
 =e^{-k \tau  E_x(\frac{R}{\tau}+\frac{\mu}{2 \tau},P_{SX})}. \label{chnexp}
\end{align}
Next, consider the type II error event that happens when an error occurs at the channel decoder. We need to consider two separate cases: $I(U;W)+\mu>R$ and $I(U;W )+\mu \leq R$. Note that in the former case, binning is performed and type II error happens at the decoder only if a sequence $W^k(l)$ exists in the wrong bin $\hat M \neq M=f_B(J)$ such that $(V^k,W^k(l)) \in T_{[VW]_{\tilde \delta}}^k$. 
As noted in \cite{Lim-minero-kim-2015}, the calculation of the probability of this event does not follow from the standard random coding argument usually encountered in achievability proofs due to the fact that the chosen codeword $W^k(J)$ depends on the entire codebook. 
Following steps similar to those in \cite{Lim-minero-kim-2015}, we analyze the probability of this event (averaged over codebooks $\mathcal{C}$ and random binning) as follows.
We first consider the case when $I(U;W)+\mu>R$. 
\begin{flalign}
&\mathbb{P}(\mathcal{D}_0|\mathcal{F}_1) \leq  \mathbb{P}(~\exists~ W^k(l):~ f_B(l)=\hat M \neq 1,~(W^k(l),v^k) \in T_{[WV]_{\tilde\delta}}^k|\mathcal{F}_1) \notag \\
& \leq \sum_{l=2}^{M_k'} \sum_{ \hat m \neq 1} \mathbb{P} (\hat M=\hat m| \mathcal{F}_1)~ \mathbb{P}((W^k(l),v^k) \in T_{[WV]_{\tilde \delta}}^k:~ f_B(l)=\hat m|\mathcal{F}_1) \notag \\
&=\sum_{l=2}^{M_k'} \sum_{\hat m \neq 1} \mathbb{P} (\hat M=\hat m| \mathcal{F}_1)~ \sum_{\substack{\tilde w^k:\\
(\tilde w^k,v^k)\in T_{[WV]_{\tilde \delta}}^k}} \mathbb{P}(W^k(l)=\tilde w^k:~ f_B(l)=\hat m|\mathcal{F}_1)\notag &&
\end{flalign}
\begin{flalign}
&=\sum_{l=2}^{M_k'} \sum_{\hat m \neq 1} \mathbb{P} (\hat M=\hat m| \mathcal{F}_1)~\sum_{\substack{\tilde w^k:\\
(\tilde w^k,v^k)\in T_{[WV]_{\tilde \delta}}^k}}  \mathbb{P}(W^k(l)=\tilde w^k|\mathcal{F}_1)~ \frac{1}{e^{kR}} \notag \\
&=\sum_{l=2}^{M_k'} \sum_{\substack{\tilde w^k:\\
(\tilde w^k,v^k)\in T_{[WV]_{\tilde \delta}}^k}} \mathbb{P}(W^k(l)=\tilde w^k|\mathcal{F}_1)~ \frac{1}{e^{kR}}. \label{error-exponentbndapply}&&
\end{flalign}
Let $\mathcal{C}_{1,l}^-= \mathcal{C} \backslash \{W^k(1), W^k(l)\}$. Then,
\begin{align}
&\mathbb{P}(W^k(l)=\tilde w^k|\mathcal{F}_1) = \sum_{\mathcal{C}_{1,l}^-=c} \mathbb{P}(\mathcal{C}_{1,l}^-=c|\mathcal{F}_1) \mathbb{P}(W^k(l)=\tilde w^k|\mathcal{F}_1, \mathcal{C}_{1,l}^-=c).\label{avgcodebminus}
\end{align}
 The term in  \eqref{avgcodebminus} can be upper bounded as follows:
\begin{flalign}
 & \mathbb{P}(W^k(l)=\tilde w^k|\mathcal{F}_1, ~\mathcal{C}_{1,l}^-=c) \notag \\
&=\mathbb{P}(W^k(l)=\tilde w^k| U^k=u^k,V^k=v^k, ~ \mathcal{C}_{1,l}^-=c)~ \frac{\mathbb{P}(W^k(1)=w^k| W^k(l)=\tilde w^k, U^k=u^k,V^k=v^k, \mathcal{C}_{1,l}^-=c)}{\mathbb{P}(W^k(1)=w^k|  U^k=u^k,V^k=v^k, \mathcal{C}_{1,l}^-=c)} \notag  \\
& \qquad \frac{\mathbb{P}(J=1| W^k(1)=w^k, W^k(l)=\tilde w^k, U^k=u^k,V^k=v^k, \mathcal{C}_{1,l}^-=c)}{\mathbb{P}(J=1| W^k(1)=w^k, U^k=u^k,V^k=v^k, \mathcal{C}_{1,l}^-=c)}  \label{boundcdwrd} \\
&\qquad \frac{\mathbb{P}(f_B(J)=1| J=1, W^k(1)=w^k, W^k(l)=\tilde w^k, U^k=u^k,V^k=v^k,~\mathcal{C}_{1,l}^-=c)}{\mathbb{P}(f_B(J)=1| J=1, W^k(1)=w^k, U^k=u^k,V^k=v^k, \mathcal{C}_{1,l}^-=c)} \notag \\
&\qquad \frac{\mathbb{P}(\mathcal{E}_{NE},~ \mathcal{E}_{CE}|f_B(J)=1, J=1, W^k(1)=w^k, W^k(l)=\tilde w^k, U^k=u^k,V^k=v^k,~\mathcal{C}_{1,l}^-=c)}{\mathbb{P}(\mathcal{E}_{NE},~ \mathcal{E}_{CE}|f_B(J)=1,J=1, W^k(1)=w^k, U^k=u^k,V^k=v^k, \mathcal{C}_{1,l}^-=c)}. \notag &&
\end{flalign}

Since the codewords are generated independently of each other and the binning operation is independent of the codebook generation, we have 
\begin{flalign*}
&\mathbb{P}(W^k(1)=w^k| W^k(l)=\tilde w^k, U^k=u^k,V^k=v^k, \mathcal{C}_{1,l}^-=c)  =\mathbb{P}(W^k(1)=w^k|  U^k=u^k,V^k=v^k, \mathcal{C}_{1,l}^-=c), &&
\end{flalign*}
and 
\begin{flalign*}
&\mathbb{P}(f_B(J)=1| J=1, W^k(1)=w^k, W^k(l)=\tilde w^k, U^k=u^k,V^k=v^k,  \mathcal{C}_{1,l}^-=c) \notag \\
& =\mathbb{P}(f_B(J)=1| J=1, W^k(1)=w^k, U^k=u^k,V^k=v^k, \mathcal{C}_{1,l}^-=c).&&
\end{flalign*}
Also, note that
\begin{flalign}
 &\mathbb{P}(\mathcal{E}_{NE},~ \mathcal{E}_{CE}|f_B(J)=1, J=1, W^k(1)=w^k, W^k(l)=\tilde w^k, U^k=u^k,V^k=v^k,~\mathcal{C}_{1,l}^-=c)   \notag \\
 &=\mathbb{P}(\mathcal{E}_{NE},~ \mathcal{E}_{CE}|f_B(J)=1,J=1, W^k(1)=w^k, U^k=u^k,V^k=v^k, \mathcal{C}_{1,l}^-=c). \notag &&
\end{flalign}
Next, consider the term in \eqref{boundcdwrd}.  Let $N(u^k, \mathcal{C}_{1,l}^-)= \vert\{ w^k(l') \in \mathcal{C}_{1,l}^-:l' \neq 1,~ l' \neq l,~ (w^k(l'),u^k) \in T_{[WU]_{\delta}}^k \}\vert$. Recall that if there are multiple sequences in codebook $\mathcal{C}$ that are jointly typical with the observed sequence $U^k$, then the encoder selects one of them uniformly at random. Also, note that given $\mathcal{F}_1$, $(w^k, u^k) \in T_{[WU]_{\delta}}^k$. Thus, if $(\tilde w^k, u^k) \in T_{[WU]_{\delta}}^k$, then

\begin{flalign*}
&\frac{\mathbb{P}(J=1| W^k(1)=w^k, W^k(l)=\tilde w^k, U^k=u^k,V^k=v^k,\mathcal{E}_{NE},~ \mathcal{E}_{CE}, ~ \mathcal{C}_{1,l}^-=c)}{\mathbb{P}(J=1| W^k(1)=w^k, U^k=u^k,V^k=v^k, \mathcal{C}_{1,l}^-=c)} \notag \\
&=\left[ \frac{1}{N(u^k, \mathcal{C}_{1,l}^-)+2} \right]\frac{1}{\mathbb{P}(J=1| W^k(1)=w^k, U^k=u^k,V^k=v^k, \mathcal{C}_{1,l}^-=c)}  \\
&\leq \frac{N(u^k, \mathcal{C}_{1,l}^-)+2}{N(u^k, \mathcal{C}_{1,l}^-)+2}  = 1. &&
\end{flalign*}
If $(\tilde w^k, u^k) \notin T_{[WU]_{\delta}}^k$, then 
\begin{flalign*}
&\frac{\mathbb{P}(J=1| W^k(1)=w^k, W^k(l)=\tilde w^k, U^k=u^k,V^k=v^k, \mathcal{C}_{1,l}^-=c)}{\mathbb{P}(J=1| W^k(1)=w^k, U^k=u^k,V^k=v^k, \mathcal{C}_{1,l}^-=c)} \notag \\
&=\left[ \frac{1}{N(u^k, \mathcal{C}_{1,l}^-)+1} \right]\frac{1}{\mathbb{P}(J=1| W^k(1)=w^k, U^k=u^k,V^k=v^k, \mathcal{C}_{1,l}^-=c)}  \notag \\
&\leq \frac{N(u^k, \mathcal{C}_{1,l}^-)+2}{N(u^k, \mathcal{C}_{1,l}^-)+1}  \leq 2. &&
\end{flalign*}
Hence, the term in \eqref{avgcodebminus} can be upper bounded as
\begin{flalign}
&\mathbb{P}(W^k(l)=\tilde w^k|\mathcal{F}_1)  \notag \\
& \leq  \sum_{\mathcal{C}_{1,l}^-=c} \mathbb{P}(\mathcal{C}_{1,l}^-=c|\mathcal{F}_1) ~2~ \mathbb{P}(W^k(l)=\tilde w^k| U^k=u^k,V^k=v^k, ~ \mathcal{C}_{1,l}^-=c) \notag \\
& = 2~ \mathbb{P}(W^k(l)=\tilde w^k| U^k=u^k,V^k=v^k)=2~ \mathbb{P}(W^k(l)=\tilde w^k). \label{bndtwiceprob} &&
\end{flalign}
 Substituting \eqref{bndtwiceprob} in \eqref{error-exponentbndapply}, we obtain
 \begin{flalign}
 \mathbb{P}(\mathcal{D}_0|\mathcal{F}_1) & \leq \sum_{l=1}^{M_k'} \sum_{\substack{\tilde w^k:\\
(\tilde w^k,v^k)\in T_{[WV]_{\tilde \delta}}^k}} 2 ~ \mathbb{P}(W^k(l)=\tilde w^k) ~ \frac{1}{e^{kR}} \notag \\
 &=\sum_{l=1}^{M_k'} \sum_{\substack{\tilde w^k:\\
(\tilde w^k,v^k)\in T_{[WV]_{\tilde \delta}}^k}} 2 \cdot e^{-k (H(W)- O(\hat \delta))} ~ \frac{1}{e^{kR}} \notag \\
&=2~ M_k' ~e^{k(H(W|V)+ \delta)}~ e^{-k (H(W)- O(\hat \delta))} ~ \frac{1}{e^{kR}} \notag \\
 & \leq e^{-k(R-I(U;W|V)-\delta_2^{(k)})}, \label{expnchnerrbin}&&
 \end{flalign}
 where $\delta_2^{(k)} := O(\delta)+ \frac {\log(2)}{k}$.
For the case $I(U;W)+\mu\leq R$ (when binning is not done), the terms can be bounded similarly using \eqref{bndtwiceprob} as follows. 
\begin{flalign}
\mathbb{P}(\mathcal{D}_0|\mathcal{F}_1) &=\sum_{\hat m \neq 1 } \mathbb{P}(\hat M=\hat m|\mathcal{F}_1) ~\mathbb{P}((W^k(\hat m),v^k) \in T_{[WV]_{\tilde \delta}}^k|\mathcal{F}_1) \notag &&
\end{flalign}
\begin{flalign}
& \leq \sum_{\hat m \neq 1 } \mathbb{P}(\hat M=\hat m|\mathcal{F}_1) ~ \sum_{\substack{\tilde w^k:\\
(\tilde w^k,v^k)\in T_{[WV]_{\tilde \delta}}^k}} 2~ \mathbb{P}(W^k( \hat m)=\tilde w^k)\notag \\
&\leq e^{-k(I(V;W)- \delta_2^{(k)})}.\label{expchnerrnobin} &&
\end{flalign}

Next, consider the event when there are no  encoding or channel errors, i.e., $\mathcal{E}_{NE} ~\cap ~\mathcal{E}_{CE}^c$. 
For the case $I(U;W)+\mu>R$, the binning error event denoted by $\mathcal{E}_{BE}$  happens when a wrong codeword $W^k(l),~ l \neq J$, is retrieved from the bin with index $M$ by the empirical entropy decoder such that $(W^k(l), V^k) \in T_{[WV]_{\delta}}^k$. Let $P_{\tilde U\tilde V\tilde W}$ denote the type of $P_{U^kV^kW^k(J)}$. 
Note that $P_{\tilde U \tilde W} \in \mathcal{T}_{[UW]_{\delta}}^k $ when $\mathcal{E}_{NE}$ holds. If $H(\tilde W| \tilde V)< H(W|V)$, then in the bin with index $M$, there exists a codeword with empirical entropy strictly less than $H(W|V)$. Hence, the decoded codeword $\hat W^k $ is such that $(\hat W^k,V^k)\notin T_{[WV]_{\tilde \delta}}^k$(asymptotically) since  $(\hat W^k,~V^k)\in T_{[WV]_{\tilde \delta}}^k$ necessarily implies that $H_e( \hat W^k| V^k) \geq H(W|V)-O(\delta)$ (for $\delta$ small enough). 
Consequently, a type II error can happen under the event $\mathcal{E}_{BE}$ only when $H(\tilde W| \tilde V) \geq  H(W|V)-O(\delta)$. The probability of the event $\mathcal{E}_{BE}$ can be upper bounded under this condition as follows: 
\begin{flalign}
&\mathbb{P}(\mathcal{E}_{BE}|\mathcal{F}_2) \notag\\ 
& \leq \mathbb{P} \left( \exists ~l \neq 1, ~l \in [M_k'] :~f_B(l)=1\mbox{ and }(W^k(l),v^k) \in T_{[WV]_{\tilde \delta}}^k| \mathcal{F}_2 \right) \notag \\
& \leq \sum_{l=2}^{M_k'}\mathbb{P} \left((W^k(l),v^k) \in T_{[WV]_{\tilde \delta}}^k| \mathcal{F}_2 \right)~ \mathbb{P} \left( f_B(l)=1|\mathcal{F}_2,(W^k(l),v^k) \in T_{[WV]_{\tilde \delta}}^k\right)\notag \\
&=  \sum_{l=2}^{M_k'}\mathbb{P} \left((W^k(l),v^k) \in T_{[WV]_{\tilde \delta}}^k| \mathcal{F}_2 \right)~  e^{-kR} \notag \\
 & \leq \sum_{l=2}^{M_k'}\sum_{\substack{\tilde w^k:\\
(\tilde w^k,v^k)\in T_{[WV]_{\tilde \delta}}^k}} 2~ \mathbb{P}(W^k( l)=\tilde w^k)~  e^{-kR} \label{bndtwicenchnerr} \\
&=e^{-k(R-I(U;W|V)- \delta_2^{(k)})}. \label{expnchnerrbinerr} &&
\end{flalign}
 In \eqref{bndtwicenchnerr}, we used the fact that
\begin{align}
\mathbb{P} \left(W^k(l)= \tilde w^k| \mathcal{F}_2 \right) \leq 2~ \mathbb{P}(W^k(l)=\tilde w^k),
\end{align}
which follows in a similar way as \eqref{bndtwiceprob}.
Also, note that, by definition,
$\mathbb{P}(\mathcal{D}_0|\mathcal{F}_{21})=1$. 

We proceed to analyze the R.H.S of \eqref{finsplitfirst} which upper bounds the type II error probability.
Towards this end, we first focus on the the case when $\mathcal{E}_{NE}$ holds.
From \eqref{lsttermexp}, it follows that
\begin{flalign}
&\sum_{(u^k,v^k) \in \Ucal^k \times \V^k} \mathbb{P}(U^k=u^k,V^k=v^k|H=1) ~\mathbb{P}(\mathcal{D}_0|U^k=u^k,V^k=v^k,\mathcal{E}_{NE}) &&
\end{flalign}
\begin{flalign}
&= \sum_{(u^k,v^k) \in \Ucal^k \times \V^k} \mathbb{P}(U^k=u^k,V^k=v^k|H=1)~ \mathbb{P}(\mathcal{D}_0|U^k=u^k,V^k=v^k,J=1, f_B(J)=1,\mathcal{E}_{NE}). \label{firsttermanlys}&&
\end{flalign}
Rewriting the summation in \eqref{firsttermanlys} as the sum over the types and  sequences within a type, we obtain
\begin{flalign}
&\mathbb{P}(\mathcal{D}_0|~\mathcal{E}_{NE},H=1) \notag \\
&= \sum_{ \substack{P_{\tilde U \tilde V \tilde W} \\ \in   \mathcal{T}^k_{\Ucal \V  \W}} }\sum_{\substack{(u^k,v^k,w^k) \\ \in T_{P_{\tilde U \tilde V \tilde W}}}} \Big[\mathbb{P}(U^k=u^k,V^k=v^k|H=1)~\mathbb{P}(\mathcal{D}_0|\mathcal{F}) \notag \\ &\qquad \qquad  \qquad \qquad \qquad\mathbb{P}(W^k(1)=w^k|U^k=u^k,V^k=v^k, J=1, f_B(J)=1, \mathcal{E}_{NE}) \Big].\label{t2rewritsumm} &&
\end{flalign}
We also have 
\begin{flalign}
&\mathbb{P}(U^k=u^k,V^k=v^k|H=1) ~\mathbb{P}(W^k(1)=w^k|U^k=u^k,V^k=v^k, J=1, f_B(J)=1, \mathcal{E}_{NE})  \notag \\
&= \left[\prod_{i=1}^k Q_{UV}(u_i,v_i) \right] \mathbb{P}(W^k(1)=w^k|U^k=u^k,V^k=v^k,J=1, f_B(J)=1,\mathcal{E}_{NE}) \notag \\
& \leq \left[\prod_{i=1}^k Q_{UV}(u_i,v_i) \right] \frac{1}{|T_{P_{\tilde W|\tilde U}}|} \leq  e^{-k(H(\tilde U \tilde V)+D(P_{\tilde U \tilde V}||Q_{UV})+H(\tilde W|\tilde U)-\frac{1}{k}|\Ucal||\W|\log(k+1))},\label{firsteqexp} &&
\end{flalign}
where $P_{\tilde U \tilde V \tilde W}$ denotes the type of the sequence $(u^k,v^k,w^k)$.

With \eqref{chnexp}, \eqref{expnchnerrbin}, \eqref{expchnerrnobin}, \eqref{expnchnerrbinerr} and  \eqref{firsteqexp}, we have the necessary machinery to analyze \eqref{t2rewritsumm}. First, consider that the event $\mathcal{E}_{NE} \cap \mathcal{E}_{CE}^c \cap \mathcal{E}_{BE}^c$ holds. In this case,
\begin{flalign} \label{condt2errnoerr}
\mathbb{P}(\mathcal{D}_0|\mathcal{F}_{22}) &=\mathbb{P}(\mathcal{D}_0|U^k=u^k,V^k=v^k,J=1, f_B(J)=1, W^k(1)=w^k,\mathcal{E}_{NE}, \mathcal{E}_{CE}^c,\mathcal{E}_{BE}^c) \notag \\
&=\begin{cases}
 1,  \mbox{ if }P_{u^k w^k}  \in T_{[UW]_{\delta}}^k \mbox{ and } P_{v^k w^k}  \in T_{[VW]_{\tilde \delta}}^k,  \\
0, \mbox{ otherwise}. 
 \end{cases} &&
\end{flalign}
Thus, the  following terms in \eqref{t2rewritsumm} can be simplified (for sufficiently large $k$) as follows:
\begin{flalign}
 & \sum_{\substack{P_{\tilde U \tilde V \tilde W} \\ \in  \mathcal{T}^k_{\Ucal \V \W}}}\sum_{\substack{(u^k,v^k,w^k) \\ \in T_{P_{\tilde U \tilde V \tilde W}}}} \Big[ \mathbb{P}(U^k=u^k,V^k=v^k|H=1) ~ \mathbb{P}(\mathcal{E}_{CE}^c|\mathcal{F}) ~ \mathbb{P}(\mathcal{E}_{BE}^c|\mathcal{F}_2)  ~\mathbb{P}(\mathcal{D}_0|\mathcal{F}_{22})  \notag \\
 & \qquad \qquad \qquad \qquad \qquad \qquad \qquad  \mathbb{P}(W^k(1)=w^k|U^k=u^k,V^k=v^k, J=1, f_B(J)=1, \mathcal{E}_{NE}) \Big] \notag \\
  & \leq \sum_{\substack{P_{\tilde U \tilde V \tilde W} \\ \in   \mathcal{T}^k_{\Ucal \V \W}}}\sum_{\substack{(u^k,v^k,w^k) \\ \in T_{P_{\tilde U \tilde V \tilde W}}}} \Big[ \mathbb{P}(U^k=u^k,V^k=v^k|H=1) ~\mathbb{P}(\mathcal{D}_0|\mathcal{F}_{22}) \notag \\
 & \qquad \qquad \qquad \qquad \qquad \qquad \quad~~ ~\mathbb{P}(W^k(1)=w^k|U^k=u^k,V^k=v^k, J=1, f_B(J)=1, \mathcal{E}_{NE})  \Big] \notag \\
 & \leq (k+1)^{|\Ucal||\V||\W|}\max_{\substack{ P_{\tilde U \tilde V \tilde W} \in \\  \hat{\mathcal{T}}_1^{(k)}(P_{UW}, P_{VW}) } } e^{kH(\tilde U \tilde V \tilde W)} e^{-k(H(\tilde U \tilde V)+D(P_{\tilde U \tilde V}||Q_{UV})+H(\tilde W|\tilde U)-\frac{1}{k}|\Ucal||\W|\log(k+1))} \notag \\ &= e^{-k\tilde E_{1k}}, \label{expE1} &&
    \end{flalign}
where,
\begin{flalign}
   &  \hat{\mathcal{T}}_1^{(k)}(P_{UW}, P_{VW}) := \{ P_{\tilde U \tilde V \tilde W }: P_{\tilde U \tilde W } \in T_{[UW]_{\delta}}^k \mbox{ and } P_{\tilde V \tilde W } \in T_{[VW]_{\tilde{\delta}}}^k \}, &&
   \end{flalign}
\begin{flalign}
\mbox{and }&\tilde E_{1k} :=  \min_{\substack{ P_{\tilde U \tilde V \tilde W}~ \in \notag \\ \hat{\mathcal{T}}_1^{(k)}(P_{UW}, P_{VW}) } } H(\tilde U \tilde V)+D(P_{\tilde U \tilde V}||Q_{UV})+H(\tilde W|\tilde U)-H(\tilde U \tilde V \tilde W)-\frac{1}{k}|\Ucal||\V||\W|\log(k+1) \notag \\
&\qquad \qquad \qquad \qquad \qquad  -\frac{1}{k}|\Ucal||\W|\log(k+1). \label{nonasympt1set} &&
\end{flalign}
To obtain \eqref{expE1}, we used \eqref{firsteqexp} and \eqref{condt2errnoerr}. 
Note that for $\delta$ small enough,
\begin{flalign}
\tilde E_{1k}  &\overset{(k)}{\geq } \min_{\substack{ P_{\tilde U \tilde V \tilde W}~  \in \\  {\mathcal{T}}_1(P_{UW}, P_{VW}) } } \sum P_{\tilde U \tilde V \tilde W} \log \left(\frac{P_{\tilde U \tilde V}}{Q_{UV}} \frac{1}{P_{\tilde U \tilde V}} \frac{P_{\tilde U}}{P_{\tilde U \tilde W}} P_{\tilde U \tilde V \tilde W}\right)-O(\delta) \notag \\
& =\min_{\substack{ P_{\tilde U \tilde V \tilde W}~  \in \\  \mathcal{T}_1(P_{UW}, P_{VW}) } } D(P_{\tilde U \tilde V \tilde W}|| Q_{UVW})-O(\delta) =E_1(P_{W|U})-O(\delta), \label{fineqexp} &&
\end{flalign}

Next, consider the terms corresponding to the event $\mathcal{E}_{NE} \cap \mathcal{E}_{CE}^c \cap \mathcal{E}_{BE}$ in \eqref{t2rewritsumm}. Note that given the event $\mathcal{F}_{21}= \{U^k=u^k,V^k=v^k,J=1,f_B(J)=1,W^k(1)=w^k,~\mathcal{E}_{NE}, ~\mathcal{E}_{CE}^c,~\mathcal{E}_{BE}\}$ occurs, $P_{u^k w^k} \in T_{[UW]_{\delta}}^k$. Also, $\mathcal{D}_0$ can  happen only if $H_e(w^k|v^k) \geq H(W|V)-O(\tilde \delta)$, and $P_{v^k} \in T_{[V]_{\check{\delta}}}^k$. 
Using these facts to simplify the terms corresponding to the event $\mathcal{E}_{NE} \cap \mathcal{E}_{CE}^c \cap \mathcal{E}_{BE}$ in \eqref{t2rewritsumm}, we obtain
\begin{flalign}
 & \sum_{\substack{P_{\tilde U \tilde V \tilde W} \\ \in  \mathcal{T}^k_{\Ucal \V  \W}}}\sum_{\substack{(u^k,v^k,w^k) \\ \in T_{P_{\tilde U \tilde V \tilde W}}}} \Big[ \mathbb{P}(U^k=u^k,V^k=v^k|H=1) ~ \mathbb{P}(\mathcal{E}_{CE}^c|\mathcal{F}) ~ \mathbb{P}(\mathcal{E}_{BE}|\mathcal{F}_2)  ~\mathbb{P}(\mathcal{D}_0|\mathcal{F}_{21}) \notag \\
 & \qquad \qquad \qquad \qquad  \qquad  \qquad  \quad ~ \mathbb{P}(W^k(1)=w^k|U^k=u^k,V^k=v^k, J=1, f_B(J)=1, \mathcal{E}_{NE}) \Big] \notag \\
  & \leq \sum_{\substack{P_{\tilde U \tilde V \tilde W} \\\in  \mathcal{T}^k_{\Ucal \V  \W}}}\sum_{\substack{(u^k,v^k,w^k) \\ \in T_{P_{\tilde U \tilde V \tilde W}}}} \Big[ \mathbb{P}(U^k=u^k,V^k=v^k|H=1) ~\mathbb{P}(\mathcal{E}_{BE}|\mathcal{F}_2)  ~\mathbb{P}(\mathcal{D}_0|\mathcal{F}_{21}) \notag \\
 & \qquad \qquad \qquad  \qquad \qquad \qquad \quad    ~ \mathbb{P}(W^k(1)=w^k|U^k=u^k,V^k=v^k, J=1, f_B(J)=1, \mathcal{E}_{NE}) \Big] \notag \\[5 pt]
 & \leq   \max_{\substack{ P_{\tilde U \tilde V \tilde W} \in \notag \\  \hat{\mathcal{T}}_2^{(k)}(P_{UW}, P_{V}) } } e^{kH(\tilde U \tilde V \tilde W)} e^{-k\left(H(\tilde U \tilde V)+D(P_{\tilde U \tilde V}||Q_{UV})+H(\tilde W|\tilde U)+R-I(U;W|V)-O(\delta) \right) } \notag \\
 & \qquad \qquad \qquad \qquad   e^{\left(|\Ucal||\V||\W|\log(k+1)+|\Ucal||\W|\log(k+1)\right)} \notag \\
& =   e^{-k\tilde E_{2k}},&&
\end{flalign}
where,
\begin{flalign}
    &\hat{\mathcal{T}}_2^{(k)}(P_{UW}, P_{V}):=\{ P_{\tilde U \tilde V \tilde W }: P_{\tilde U \tilde W } \in T_{[UW]_{\delta}}^k,  P_{\tilde V } \in T_{[V]_{\check{\delta}}}^k  \mbox{ and } H(\tilde W| \tilde V) \geq H(W|V)-O(\delta) \}, \label{nonsymptsett2}&& 
        \end{flalign}
        and
    \begin{flalign}
 &\tilde{E}_{2k} :=  \min_{\substack{ P_{\tilde U \tilde V \tilde W} \in \notag \\  {\mathcal{T}}_2(P_{UW}, P_{V}) } } H(\tilde U \tilde V)+D(P_{\tilde U \tilde V}||Q_{UV})+H(\tilde W|\tilde U)+R-I(U;W|V)-\frac{1}{k}|\Ucal||\V||\W|\log(k+1) \notag &&
     \end{flalign} 
         \begin{flalign}
 &\qquad \qquad \qquad  \qquad -\frac{1}{k}|\Ucal||\W|\log(k+1)-O(\delta) \notag \\
&\qquad \overset{(k)}{\geq } E_2(P_{W|U},P_{SX},R)-O(\delta). \label{e2ktendstoe2} &&
\end{flalign}
Also, note that $\mathcal{E}_{BE}$ occurs only when $I(U;W)+\mu>R$.

Next, consider that the event $\mathcal{E}_{NE} \cap \mathcal{E}_{CE}$ holds. As in the case above, note that given $\mathcal{F}_1= \{U^k=u^k,V^k=v^k,J=1,f_B(J)=1,W^k(1)=w^k,~\mathcal{E}_{NE},~\mathcal{E}_{CE}\}$, $P_{u^k w^k} \in T_{[UW]_{\delta}}^k$ and $\mathcal{D}_0$ occurs only if $P_{v^k} \in T_{[V]_{\check{\delta}}}^k$. Using these facts and eqns. \eqref{expnchnerrbin}, \eqref{expchnerrnobin} and \eqref{chnexp}, it can be shown  that the terms corresponding to this event in \eqref{t2rewritsumm} results in the factor   $E_3(P_{W|U},P_{SX},R,\tau)-O(\delta)$ in the error-exponent.

 Finally, we analyze the case when the event $\mathcal{E}_{NE}^c$ occurs. Since the encoder declares $H_1$ if $\hat M=0$, it is clear that $\mathcal{D}_0$ occurs only when the channel error event $\mathcal{E}_{CE}$ happens.
 Thus, we have
 \begin{flalign}
\mathbb{P}(\mathcal{D}_0|~U^k=u^k,V^k=v^k,~\mathcal{E}_{NE}^c)  =& \mathbb{P}(\mathcal{E}_{CE} |~U^k=u^k,V^k=v^k,~\mathcal{E}_{NE}^c) \notag\\
&\qquad \mathbb{P}(\mathcal{D}_{0} |~U^k=u^k,V^k=v^k,~\mathcal{E}_{NE}^c \cap  ~\mathcal{E}_{CE}).  \label{spclerror} 
  \end{flalign}
It follows from Borade et al.'s  coding scheme \cite{Borade-09} that asymptotically,
  \begin{flalign}
\mathbb{P}(\mathcal{E}_{CE} |~U^k=u^k,V^k=v^k,~\mathcal{E}_{NE}^c) \leq e^{-n(E_m\left(P_{SX}\right)-O(\delta))}=e^{-k \tau (E_m\left(P_{SX}\right)-O(\delta))}. \label{spclexp} 
 \end{flalign}
 When binning is performed at the encoder, $\mathcal{D}_0$ occurs only if there exists a sequence $\hat W^k$ in the bin $\hat M \neq 0$ such that $(\hat W^k, V^k) \in T_{[WV]_{\tilde \delta}}^k$. 
 Also, recalling  that the encoder sends the error message $M=0$ independent of the source codebook $\mathcal{C}$, it can be shown using standard arguments that for such $v^k \in T_{[V]_{\check{\delta}}}^k$,
 \begin{flalign}
 \mathbb{P}(\mathcal{D}_{0} |~U^k=u^k,V^k=v^k,~\mathcal{E}_{NE}^c \cap  ~\mathcal{E}_{CE}) \leq e^{-k(R-I(U;W|V)-O(\delta))}. \label{spclquanterr} 
 \end{flalign}
 Thus, from \eqref{spclerror}, \eqref{spclexp} and  \eqref{spclquanterr}, we obtain (asymptotically) that,
 \begin{flalign}
& \sum_{u^k,v^k}\mathbb{P}(U^k=u^k,V^k=v^k|H=1) ~\mathbb{P}(\mathcal{D}_{0} |~U^k=u^k,V^k=v^k,~\mathcal{E}_{NE}^c \cap  ~\mathcal{E}_{CE}) \notag \\
&\leq  e^{-k(R-I(U;W|V)+D(P_V||Q_V)+\tau E_m\left(P_{SX}\right)-O(\delta))}.&& \label{binnspclerrmsg}
\end{flalign}
  On the other hand, when binning is not performed, $\mathcal{D}_0$ occurs only if $(W^k(\hat M),V^k) \in T_{[WV]_{\tilde \delta}}^k$ and in this case, we obtain (asymptotically) that,
 \begin{flalign}
& \sum_{u^k,v^k}\mathbb{P}(U^k=u^k,V^k=v^k|H=1) ~\mathbb{P}(\mathcal{D}_{0} |~U^k=u^k,V^k=v^k,~\mathcal{E}_{NE}^c \cap  ~\mathcal{E}_{CE}) \notag \\
&\leq   e^{-k\left(I(V;W)+D(P_V||Q_V)+\tau E_m\left(P_{SX}\right)-O(\delta)\right)}. \label{spclmsgerrnobin} &&
\end{flalign}
This results in the factor $E_4(P_{W|U},P_{SX},R,\tau)-O(\delta)$ in the error-exponent. Since the error-exponent is lower bounded  by the minimal value of the exponent due to the various type II error events, the  proof of the theorem is complete by noting that $\delta>0$ is arbitrary.

  \section{Proof of Theorem \ref{thm:jhtcceed}} \label{JHTCCproof}
We only give a sketch of the proof as the intermediate steps follow similarly to those in the proof of Theorem \ref{thm:shtcceed}. We will use the random coding method combined with the expurgation technique as explained in the proof of Theorem \ref{thm:shtcceed}, to guarantee the existence of at least one deterministic codebook that achieves the type I error probability and error-exponent claimed in Theorem \ref{thm:jhtcceed}. For brevity, we will denote information theoretic quantities like $I_{\hat P}(U,S;\bar W)$, $T_{[\hat P_{US\bar W}]_{\delta}}^n$, etc., that are computed with respect to joint distribution  $ \hat P_{UVS\bar WX'XY}$ given below in \eqref{jointdisthybdef}
 by $I(U,S;\bar W)$, $T_{[US\bar W]_{\delta}}^n$, etc.

  Fix distributions $(P_S,P_{\bar {W}|US},P_{X'|US},P_{X|US\bar {W}}) \in \mathcal{B}_h$ and a positive number $\delta>0$. Let $ \mu=O(\delta)$ subject to  constraints that will be specified below. Let $\hat \delta:=|\bar{\mathcal{W}}|\delta$, $\delta' :=\frac{\delta}{2}$, $\bar \delta:= \frac{\delta'}{|\V|}$, $\tilde \delta:=2 \delta$, and 
  \begin{align}
   \hat P_{UVS\bar WX'XY}(P_S,P_{\bar{W}|US}, P_{X'|S}, P_{X|US\bar{W}}):= P_{UV}P_SP_{\bar{W}|US} P_{X'|US} P_{X|US\bar{W}}P_{Y|X}. \label{jointdisthybdef}
\end{align}
  Generate a sequence $S^n$ i.i.d. according to $\prod_{i=1}^n P_S(s_i)$.  
The realization  $S^n=s^n$ is revealed to both the encoder and detector. Generate the quantization codebook  $\mathcal{C}=\{\bar{w}^n(j),~j \in [e^{n(I(U,S;\bar{W})+ \mu)}]\}$, where each codeword $\bar{w}^n(j)$ is generated independently according to the distribution $\prod_{i=1}^n \hat P_{\bar{W}}$, where
\begin{align}
   \hat P_{\bar{W}}= \sum_{(u,s) \in \Ucal \times \mathcal{S}} P_U(u)P_S(s) P_{\bar{W}|US}(\bar{w}|u,s). \notag
\end{align}

\textit{Encoding}:
If $(u^n,s^n)$ is typical, i.e., $(u^n,s^n) \in T_{[US]_{\delta'}}^n$, the encoder first looks for a sequence ${\bar{w}}^n(j)$ such that $(u^n,s^n, {\bar{w}}^n(j)) \in T_{[USW]_{\delta}}^n$. If there exists multiple such codewords, it chooses one among them uniformly at random. The encoder transmits $X^n=x^n$ over the channel, where $X^n$ is  generated according to the distribution $\prod_{i=1}^n P_{X|US\bar{W}}(x_i|u_i,s_i,{\bar{w}}_i(j))$.
If $(u^n,s^n) \notin T_{[US]_{\delta'}}^k$ or such an index $j$ does not exist, the encoder generates the channel input $X'^n=x'^n$ randomly according to $\prod_{i=1}^n P_{X'|US}(x_i'|u_i,s_i)$.  

\textit{Decoding}: Given the side information sequence $V^n=v^n$, received sequence $Y^n=y^n$ and $s^n$, the detector first checks if $(v^n,s^n,y^n) \in T_{[VSY]_{\tilde \delta}}^n$, $\tilde \delta > \delta$. If the check is unsuccessful,  $\hat H=1$. Else, it searches for a typical sequence $ \hat {\bar{w}}^n={\bar{w}}^n(\hat j) \in T_{[\bar{W}]_{\hat\delta}}^k$,   in the codebook  such that  $$\hat j= \argmin_{{l:\bar{w}}^n(l)\in T_{[\bar{W}]_{\hat\delta}}^n }H_e({\bar{w}}^n(l)|v^n,s^n,y^n).$$ If $(v^n,s^n,y^n,\hat {\bar{w}}^n) \in T_{[VSY\bar{W}]_{\tilde \delta}}^n$, $\hat H=0$. Else, $\hat H=1$.  

 \textbf{Analysis of Type I error}: \\
 A type I error occurs only if one of the following events happen.
 \begin{flalign}
\tilde{\mathcal{E}}_{TE}&= \left\lbrace (U^n,V^n,S^n) \notin T_{[UVS]_{\bar \delta}}^n \right\rbrace \notag\\
\tilde{\mathcal{E}}_{EE}&=  \left\lbrace \nexists ~j \in  \left[e^{n(I(U,S;\bar{W})+\mu)}\right]: (U^n,S^n,\bar{W}^n(j)) \in T_{[US\bar{W}]_{\delta}}^n\right\rbrace \notag \\
\tilde{\mathcal{E}}_{ME}&= \left\lbrace (V^n,S^n,\bar{W}^n(J)) \notin T_{[VS\bar{W}]_{\tilde \delta}}^n \right\rbrace \notag \\
\tilde{\mathcal{E}}_{CE}&= \left\lbrace (V^n,S^n,\bar{W}^n(J),Y^n) \notin T_{[VS\bar{W}Y]_{\tilde \delta}}^n \right\rbrace \notag \\
\tilde{\mathcal{E}}_{DE}&=   \Bigg\{ \exists ~ l \in  \left[e^{n(I(U,S;\bar{W})+\mu)}\right],~ l\neq J,~\bar{W}^n(l)) \in T_{[\bar{W}]_{\hat  \delta}}^n,  H_e(\bar{W}^n(l)|V^n,S^n,Y^n) \leq H_e(\bar{W}^n(J)|V^n,S^n,Y^n) \Bigg\} \notag &&
\end{flalign}
By the weak law of large numbers, $\tilde{\mathcal{E}}_{TE}$ tends to $0$ asymptotically with $n$. The covering lemma guarantees that $\tilde{\mathcal{E}}_{EE} \cap \tilde{\mathcal{E}}_{TE}^c$ tends to $0$ doubly exponentially if $\mu=O(\delta)$ is chosen  appropriately. Given $\tilde{\mathcal{E}}_{EE}^c \cap \tilde{\mathcal{E}}_{TE}^c$ holds, it follows from the Markov lemma and the weak law of large numbers, respectively, that $\mathbb{P}(\tilde{\mathcal{E}}_{ME})$ and $\mathbb{P}(\tilde{\mathcal{E}}_{CE})$  tends to zero  asymptotically. 
Next, we consider the probability of the event $\tilde{\mathcal{E}}_{DE}$. Given that $\tilde{\mathcal{E}}_{CE}^c \cap \tilde{\mathcal{E}}_{ME}^c \cap \tilde{\mathcal{E}}_{EE}^c \cap \tilde{\mathcal{E}}_{TE}^c $ holds, note that $ H_e(\bar{W}^n(J)|V^n,S^n,Y^n) \overset{(n)}{\geq} H(\bar{W}|V,S,Y)-O(\delta)$. Hence, similarly to \eqref{type1errfinexpbin} in Appendix \ref{SHTCCproof}, it can be shown that 
\begin{align}
& \mathbb{P}(\tilde{\mathcal{E}}_{DE}|\tilde{\mathcal{E}}_{CE}^c \cap \tilde{\mathcal{E}}_{ME}^c \cap \tilde{\mathcal{E}}_{EE}^c \cap \tilde{\mathcal{E}}_{TE}^c) \leq e^{-n(I_{\hat P}(\bar{W};V,S,Y)-I_{\hat P}(U,S;\bar{W})-\delta_3^{(n)})}. \notag
\end{align}
where  $\delta_3^{(n)}\xrightarrow{(n)} O(\delta)$.
Hence, for $\delta>0$ small enough, the probability of the events causing  type I error tends to zero asymptotically since $ I(U;\bar{W}|S) < I(\bar{W};Y,V|S)$.

\textbf{Analysis of Type II error}: 
The analysis of the error-exponent is very similar to that of the SHTCC scheme given in Appendix \ref{SHTCCproof}. Hence, only a sketch of the proof is provided, with the differences from the proof of the SHTCC scheme highlighted.

Let 
\begin{align}
    \bar{\mathcal{E}}_0:=\{(U^n,S^n) \notin T_{[US]_{\delta'}}^n\}.
\end{align}
 Then,  the type 2 error probability can be written as
\begin{flalign}
&\beta\left(n,n, f^{(n,n)},g^{(n,n)}\right) \notag \\
&\leq \sum_{(u^n,v^n) \in \Ucal^n \times \V^n} \mathbb{P}(U^n=u^n,V^n=v^n|H=1)\Big[ \mathbb{P}(\tilde{\mathcal{E}}_{EE}\cap  \bar{\mathcal{E}}_0^c|U^n=u^n,V^n=v^n) \notag \\
&\qquad \qquad  \qquad \quad +\mathbb{P}(\mathcal{D}_0|U^n=u^n,V^n=v^n,\tilde{\mathcal{E}}_{NE}) +\mathbb{P}(\mathcal{D}_0|U^n=u^n,V^n=v^n, \bar{\mathcal{E}}_0) \Big], \label{finsplitfirsth} &&
\end{flalign}
where, $\tilde{\mathcal{E}}_{NE} := \tilde{\mathcal{E}}_{EE}^c \cap  \bar{\mathcal{E}}_0^c$. It is sufficient to restrict the analysis to the events $\tilde{\mathcal{E}}_{NE}$ and $ \bar{\mathcal{E}}_0$ that dominate the type 2 error.
Define the events
\begin{flalign}
&\tilde {\mathcal{E}}_{T2}=  \Big\{ 
       \exists ~ l \in  \left[e^{n(I(U,S;\bar{W})+\mu)}\right],~ l\neq J,~\bar{W}^n(l) \in T_{[\bar{W}]_{\hat  \delta}}^n,~(V^n,\bar{W}^n(l),S^n,Y^n) \in T_{[VS\bar{W}Y]_{\tilde\delta}}^n \Big\}, \label{ev1t2} \\
&\tilde {\mathcal{F}}= \{U^n=u^n,V^n=v^n,J=1,\bar{W}^n(1)=\bar{w}^n,S^n=s^n, Y^n=y^n, \tilde {\mathcal{E}}_{NE}\}, \label{ev1h} \\
&\tilde {\mathcal{F}}_1= \{U^n=u^n,V^n=v^n,J=1,\bar{W}^n(1)=\bar{w}^n,S^n=s^n, Y^n=y^n, \tilde {\mathcal{E}}_{NE}, \tilde {\mathcal{E}}_{T2}^c\}, \label{ev2h} \\
&\tilde{\mathcal{F}}_2= \{U^n=u^n,V^n=v^n,J=1,\bar{W}^n(1)=\bar{w}^n,S^n=s^n, Y^n=y^n, \tilde {\mathcal{E}}_{NE}, \tilde {\mathcal{E}}_{T2}\}. \label{ev3h} &&
\end{flalign}

By the symmetry of the codebook generation, encoding and decoding procedure, the term  
 $ \mathbb{P}(\mathcal{D}_0|U^n=u^n,V^n=v^n,J=j, ~\tilde{\mathcal{E}}_{NE})$ is independent of the value of $J$. Hence, w.l.o.g. assuming $J=1$, we can write
\begin{flalign}
&\mathbb{P}(\mathcal{D}_0|~U^n=u^n,V^n=v^n,\tilde{\mathcal{E}}_{NE})\notag \\
&=\sum_{j=1}^{e^{n(I(U,S;\bar{W})+\mu)}} ~\mathbb{P}(J=j |~U^n=u^n,V^n=v^n,\tilde{\mathcal{E}}_{NE}) ~ \mathbb{P}(\mathcal{D}_0|U^n=u^n,V^n=v^n,J=1,~\tilde{\mathcal{E}}_{NE})\notag \\
&=\mathbb{P}(\mathcal{D}_0|U^n=u^n,V^n=v^n,J=1,~\tilde{\mathcal{E}}_{NE})\notag &&
\end{flalign}
\begin{flalign}
&=\sum_{\substack{(\bar{w}^n,s^n,y^n) \\ \in ~ \bar{\W}^n \times \Scal^n \times \Y^n  }} \mathbb{P}(\bar{W}^n(1)=\bar{w}^n, S^n=s^n, Y^n=y^n |U^n=u^n,V^n=v^n, J=1, ~\tilde{\mathcal{E}}_{NE}) \notag \\
& \qquad \qquad \qquad  \qquad \mathbb{P}(\mathcal{D}_0|U^n=u^n,V^n=v^n,J=1,\bar{W}^n(1)=\bar{w}^n, S^n=s^n, Y^n=y^n,~\tilde{\mathcal{E}}_{NE}) \notag \\
&=\sum_{\substack{(\bar{w}^n,s^n,y^n) \\ \in ~ \bar{\W}^n \times \Scal^n \times \Y^n  }} \mathbb{P}(\bar{W}^n(1)=\bar{w}^n, S^n=s^n, Y^n=y^n |U^n=u^n,V^n=v^n, J=1, ~\tilde{\mathcal{E}}_{NE}) ~ \mathbb{P}(\mathcal{D}_0|~\tilde {\mathcal{F}}). \label{lsttermexph} &&
\end{flalign}

The last term in \eqref{lsttermexph} can be upper bounded using the events  in \eqref{ev1h}-\eqref{ev3h} as follows.
\begin{align}
&\mathbb{P}(\mathcal{D}_0|~\tilde {\mathcal{F}}) \leq \mathbb{P}(\mathcal{D}_0|~\tilde {\mathcal{F}}_1)+  \mathbb{P}(\tilde{\mathcal{E}}_{T2}|~\tilde {\mathcal{F}})~ \mathbb{P}(\mathcal{D}_0|~\tilde {\mathcal{F}}_2).\notag 
\end{align}

We next analyze the R.H.S of \eqref{finsplitfirsth}, which upper bounds the type 2 error probability. We can write,
\begin{equation} \label{condt2errnoerrh}
\mathbb{P}(\mathcal{D}_0|\tilde{\mathcal{F}}_{1}) =
\begin{cases}
 1,  \mbox{ if }P_{u^ns^n \bar{w}^n}  \in T_{[US\bar{W}]_{\delta}}^n \mbox{ and } P_{v^n \bar{w}^ns^ny^n} \in T_{[VS\bar{W}Y]_{\tilde \delta}}^k,  \\
0, \mbox{ otherwise}. 
 \end{cases}
\end{equation}

Hence, the terms corresponding to the event $\tilde{\mathcal{F}}_{1}$ in \eqref{finsplitfirsth} can be upper bounded (in the limit $\delta, \tilde \delta \rightarrow 0$) as 
\begin{flalign}
 &\sum_{\substack{(u^n,v^n,\bar{w}^n,s^n,y^n) \\ \in ~ \Ucal^n \times \V^n \times \bar{\W}^n \times \Scal^n \times \Y^n  }} \Big[\mathbb{P}(U^n=u^n,V^n=v^n|H=1)~\mathbb{P}(\mathcal{D}_0|\tilde{\mathcal{F}}_{1})  \notag \\
&\qquad \qquad \qquad  \qquad \qquad  \qquad \qquad  \mathbb{P}(\bar{W}^n(1)=\bar{w}^n, S^n=s^n, Y^n=y^n |U^n=u^n,V^n=v^n, J=1, ~\tilde{\mathcal{E}}_{NE})   \Big] \notag \\
&\leq  \sum_{\substack{P_{\tilde U \tilde V \tilde S  \tilde W  \tilde Y } \\ \in  \mathcal{T}^n_{\Ucal \V  \bar{\W}  \Scal  \Y}}}\sum_{\substack{(u^n,v^n,\bar{w}^n,s^n,y^n) \\ \in T_{P_{\tilde U \tilde V \tilde S  \tilde W  \tilde Y}}}} \Big[\mathbb{P}(U^n=u^n,V^n=v^n|H=1)~ \mathbb{P}(\mathcal{D}_0|\tilde{\mathcal{F}}_{1}) \notag \\
& \qquad \qquad \qquad \qquad \qquad \qquad  \qquad  \mathbb{P}(S^n=s^n,\bar{W}^n(1)=\bar{w}^n|U^n=u^n, J=1, ~\tilde{\mathcal{E}}_{NE})   \notag \\
& \qquad \qquad \qquad \qquad \qquad \qquad  \qquad \mathbb{P}(Y^n=y^n |U^n=u^n,S^n=s^n, J=1,\bar{W}^n(1)=\bar{w}^n,  ~\tilde{\mathcal{E}}_{NE})\Big] \notag \\
&\leq  \sum_{\substack{P_{\tilde U \tilde V \tilde S  \tilde W  \tilde Y }\\ \in   \mathcal{T}^n_{\Ucal \V  \bar{\W}  \Scal  \Y}}}\sum_{\substack{(u^n,v^n,\bar{w}^n,s^n,y^n) \\ \in T_{P_{\tilde U \tilde V \tilde S  \tilde W  \tilde Y}}}} \Big[\mathbb{P}(\mathcal{D}_0|\tilde{\mathcal{F}}_{1}) ~ e^{-n\left(H(\tilde U \tilde V)+D(P_{\tilde U \tilde V}||Q_{UV})\right)}  \notag \\
& \qquad \qquad \qquad \qquad  \qquad \qquad  e^{-n \left(H(\tilde S \tilde W| \tilde U)-\frac{1}{n}|\Ucal||\bar{\W}||\Scal|\log(n+1)\right)}~ e^{-n\left(H( \tilde Y|\tilde U \tilde S \tilde W)+ D(P_{\tilde Y|\tilde U \tilde S \tilde W}||\hat P_{Y|US\bar{W}}|P_{\tilde U \tilde S \tilde W})\right)}  \Big] \notag \\
&\leq \max_{\substack{ P_{\tilde U \tilde V \tilde S  \tilde W  \tilde Y} \in \\  \mathcal{T}_1'^{(n)}(\hat P_{US\bar{W}}, \hat P_{VS\bar{W}Y}) } } \Big[ e^{-n\left(H(\tilde U \tilde V)+D(P_{\tilde U \tilde V}||Q_{UV})\right)} ~e^{-n \left(H(\tilde S \tilde W| \tilde U)-\frac{1}{n}|\Ucal||\bar{\W}||\Scal|\log(n+1)\right)} \notag \\
&\qquad \qquad \qquad \qquad \qquad e^{-n\left(H( \tilde Y|\tilde U \tilde S \tilde W)+ D(P_{\tilde Y|\tilde U \tilde S \tilde W}||\hat P_{Y|US\bar{W}}|P_{\tilde U \tilde S \tilde W})\right)} e^{n\left(H(\tilde U \tilde V \tilde S  \tilde W  \tilde Y)-\frac{1}{n}||\Ucal||\V||\bar{\W}||\Scal||\Y|\log(n+1)\right)} \Big]  \notag \\
&=  e^{-n E_{1n}^*}, \label{expE1h} &&
\end{flalign}
where

\begin{flalign}
&\mathcal{T}_1'^{(n)}(\hat P_{US\bar{W}}, \hat P_{VS\bar{W}Y}):= \{P_{\tilde U \tilde V  \tilde S  \tilde W \tilde Y} \in \mathcal{T}_{\Ucal \V \mathcal{S} \W  \Y} : P_{\tilde U \tilde S \tilde W} \in T^n_{[US\bar W]_{\delta}},~ P_{ \tilde V \tilde S  \tilde W  \tilde Y} \in T^n_{[VS\bar WY]_{\tilde \delta}} \}, \notag &&
\end{flalign}
and
\begin{flalign}
&E_{1n}^*:=\min_{\substack{ P_{\tilde U \tilde V \tilde S  \tilde W  \tilde Y} \in \\  \mathcal{T}_1'(\hat P_{US\bar{W}}, \hat P_{VS\bar{W}Y}) } }  \Bigg[H(\tilde U \tilde V)+D(P_{\tilde U \tilde V}||Q_{UV})+H( \tilde S \tilde W|\tilde U)+H( \tilde Y|\tilde U  \tilde S \tilde W)- H(\tilde U \tilde V\tilde W \tilde S\tilde Y )\notag \\
&\qquad \qquad \qquad \quad  + D(P_{\tilde Y|\tilde U  \tilde S \tilde W}||\hat P_{Y|US\bar{W}}|P_{\tilde U  \tilde S \tilde W})-\frac{1}{n}(|\Ucal||\bar{\W}|+|\Ucal||\V||\bar{\W}||\Scal||\Y|)\log(n+1)\Bigg] \notag \\
&  \overset{(n)}{\geq }\min_{\substack{ P_{\tilde U \tilde V \tilde S  \tilde W  \tilde Y} \in \\  \mathcal{T}_1'(\hat P_{US\bar{W}}, \hat P_{VS\bar{W}Y}) } }\Bigg[\sum_{\tilde U \tilde V \tilde S  \tilde W  \tilde Y} P_{\tilde U \tilde V \tilde S  \tilde W  \tilde Y} \log\left(\frac{1}{P_{\tilde U \tilde V}} \frac{P_{\tilde U \tilde V}}{Q_{ U V}} \frac{P_{\tilde U}}{P_{\tilde U \tilde S \tilde W}}\frac{1}{P_{ \tilde Y| \tilde U  \tilde S \tilde W}}\frac{P_{ \tilde Y| \tilde U  \tilde S \tilde W}}{\hat P_{Y|U S \bar{W}}} P_{\tilde U \tilde V \tilde S  \tilde W  \tilde Y} \right)-O(\delta) \Bigg]\notag \\
&=\min_{\substack{ P_{\tilde U \tilde V \tilde S  \tilde W  \tilde Y} \in \\  \mathcal{T}_1'(\hat P_{US\bar{W}}, \hat P_{VS\bar{W}Y}) } }\Bigg[ D(P_{\tilde U \tilde V \tilde S  \tilde W  \tilde Y}|Q_{UV}P_{\tilde S \tilde W| \tilde U }\hat P_{Y|U S \bar{W}})-O(\delta) \Bigg]\notag \\
&=  E_1'(P_S,P_{\bar{W}|US}, P_{X|US\bar{W}})-O(\delta). \label{exp1tilde}&&
\end{flalign}
Here, \eqref{exp1tilde} follows from the fact that $P_{\tilde S \tilde W|\tilde U } \rightarrow P_{S\bar{W}|U}$ given $\tilde{\mathcal{E}}_{NE}$, as $\delta \rightarrow 0$.

Next, consider the terms corresponding to the event $\tilde{\mathcal{F}}_{2}$ in \eqref{finsplitfirsth}. Given $\tilde{\mathcal{F}}_{2}$, $P_{\tilde U \tilde S \tilde W} \in T_{[US\bar{W}]_{\delta}}^n$ and $\mathcal{D}_0$ occurs only if  $(V^n,S^n,Y^n) \in T_{[VSY]_{\delta''}}^n$, $\delta'' = |\bar{\W}| \tilde \delta$, and $H(\tilde W| \tilde V, \tilde S, \tilde Y) \geq H(\bar{W}|V,S,Y)-O(\tilde \delta)$. Thus, we have, 
\begin{flalign}
& \sum_{\substack{(u^n,v^n,\bar{w}^n,s^n,y^n) \\ \in ~ \Ucal^n \times \V^n \times \bar{\W}^n \times \Scal^n \times \Y^n  }} \Big[\mathbb{P}(U^n=u^n,V^n=v^n|H=1)~\mathbb{P}(\mathcal{D}_0|\tilde{\mathcal{F}}_{2}) ~\mathbb{P}(\tilde{\mathcal{E}}_{T2}|\tilde{\mathcal{F}}) \notag \\
& \qquad \qquad \qquad \qquad \qquad \qquad \qquad \qquad  \mathbb{P}(\bar{W}^n(1)=\bar{w}^n, S^n=s^n, Y^n=y^n |U^n=u^n,V^n=v^n, J=1, ~\tilde{\mathcal{E}}_{NE})   \Big] \notag \\
&\leq  \sum_{\substack{P_{\tilde U \tilde V \tilde S  \tilde W  \tilde Y } \in \\  \mathcal{T}^n(\Ucal \times \V \times \bar{\W} \times \Scal \times \Y)}}\sum_{\substack{(u^n,v^n,\bar{w}^n,s^n,y^n) \\ \in T_{P_{\tilde U \tilde V \tilde S  \tilde W  \tilde Y}}}} \Big[ \mathbb{P}(U^n=u^n,V^n=v^n|H=1)~ \mathbb{P}(\mathcal{D}_0|\tilde{\mathcal{F}}_{2})~ \mathbb{P}(\tilde{\mathcal{E}}_{T2}|\tilde{\mathcal{F}})  \notag \\
&~\mathbb{P}(S^n=s^n,\bar{W}^n(1)=\bar{w}^n|U^n=u^n, J=1, ~\tilde{\mathcal{E}}_{NE}) ~ \mathbb{P}(Y^n=y^n |U^n=u^n,S^n=s^n, J=1,\bar{W}^n(1)=\bar{w}^n,  ~\tilde{\mathcal{E}}_{NE}) \Big] \notag \\
&\leq \sum_{\substack{P_{\tilde U \tilde V \tilde S  \tilde W  \tilde Y } \in \\  \mathcal{T}^n(\Ucal \times \V \times \bar{\W} \times \Scal \times \Y)}}\sum_{\substack{(u^n,v^n,\bar{w}^n,s^n,y^n) \\ \in T_{P_{\tilde U \tilde V \tilde S  \tilde W  \tilde Y}}}} \Big[e^{-n\left(H(\tilde U \tilde V)+D(P_{\tilde U \tilde V}||Q_{UV})\right)}\mathbb{P}(\mathcal{D}_0|\tilde{\mathcal{F}}_{2})~ \cdot 2 \cdot e^{-n \left( I(\bar{W};V,S,Y)-I(U,S;\bar{W})-O(\delta) \right)}   \notag \\
& \qquad  \qquad  \qquad  \qquad ~~   ~e^{-n \left(H(\tilde S \tilde W| \tilde U)-\frac{1}{n}|\Ucal||\bar{\W}||\Scal|\log(n+1)\right)}~ e^{-n\left(H( \tilde Y|\tilde U \tilde S \tilde W)+ D(P_{\tilde Y|\tilde U \tilde S \tilde W}||\hat P_{Y|US\bar{W}}|P_{\tilde U \tilde S \tilde W})\right)} \Big]
  \label{bndtwicewh} \\
&\leq  \max_{\substack{ P_{\tilde U \tilde V \tilde S  \tilde W  \tilde Y} \in \\  \mathcal{T}_2'^{(n)}(\hat P_{UW}, \hat P_{VSWY}) } } \Big[ e^{-n\left(H(\tilde U \tilde V)+D(P_{\tilde U \tilde V}||Q_{UV})\right)} ~ e^{-n \left(H(\tilde S \tilde W| \tilde U)-\frac{1}{n}|\Ucal||\bar{\W}||\Scal|\log(n+1)\right)} \notag \\
&\qquad \qquad \qquad \quad  \qquad \qquad e^{-n \left( I(\bar{W};V,S,Y)-I(U,S;\bar{W})-O(\delta)-\frac {1}{n} \right)}~ e^{-n\left(H( \tilde Y|\tilde U \tilde S \tilde W)+ D(P_{\tilde Y|\tilde U \tilde S \tilde W}||\hat P_{Y|US\bar{W}}|P_{\tilde U \tilde S \tilde W})\right)} \notag \\
& \qquad \qquad \qquad \qquad \qquad ~~~ e^{n\left(H(\tilde U \tilde V \tilde S  \tilde W  \tilde Y)-\frac{1}{n}||\Ucal||\V||\bar{\W}||\Scal||\Y|\log(n+1)\right)} \Big] \notag \\  
&= e^{-n E_{2n}^*}, \label{expE2h} &&
\end{flalign}
where, 
\begin{flalign}
&\mathcal{T}_2'^{(n)}(\hat P_{US\bar{W}}, \hat P_{VS\bar{W}Y}):= \{P_{\tilde U \tilde V  \tilde S  \tilde W \tilde Y} \in \mathcal{T}_{\Ucal \V \mathcal{S} \W  \Y} : P_{\tilde U \tilde S \tilde W} \in T^n_{[US\bar W]_{\delta}},~ P_{ \tilde V \tilde S  \tilde W  \tilde Y} \in T^n_{[VS\bar WY]_{\tilde \delta}} \notag \\
& \qquad \qquad \qquad \qquad  \qquad \qquad \mbox{ and } H(\tilde W| \tilde V, \tilde S, \tilde Y) \geq H(\bar{W}|V,S,Y)-O(\delta) \}, \notag &&
\end{flalign}
and
\begin{flalign}
&E_{2n}^*\overset{(n)}{\geq }\min_{\substack{ P_{\tilde U \tilde V \tilde S  \tilde W  \tilde Y} \in \\  \mathcal{T}_2'(\hat P_{US\bar{W}}, \hat P_{VS\bar{W}Y}) } }\Bigg[ D(P_{\tilde U \tilde V \tilde S  \tilde W  \tilde Y}|Q_{UV}P_{\tilde S \tilde W| \tilde U}\hat P_{Y|US\bar{W}})+I(\bar{W};V,Y|S)-I(U;\bar{W}|S)-O(\delta) \Bigg]\notag \\
&\qquad = E_2'(P_S,P_{\bar W|US},P_{X|US\bar W})-O(\delta). \label{exp2tilde} &&
\end{flalign}
In \eqref{bndtwicewh}, we used the fact that
\begin{align}
\mathbb{P}(\tilde{\mathcal{E}}_{T2}|\tilde{\mathcal{F}}) \leq 2 \cdot e^{-n \left( I(\bar W;V,Y|S)-I(U;\bar W|S) -O(\delta)\right)}, \notag 
\end{align}
which follows from
\begin{align}
\mathbb{P} \left(\bar W^n(l)= \tilde w^n| \tilde{\mathcal{F}} \right) \leq 2~ \mathbb{P}(\bar W^n(l)=\tilde w^n). \label{bndtwiproblater} 
\end{align}
Eqn. \eqref{bndtwiproblater} can be proved similarly to \eqref{bndtwiceprob}.

Finally, we consider the case when $ \bar{\mathcal{E}}_0$ holds.
 \begin{flalign}
& \sum_{u^n,v^n}\mathbb{P}(U^n=u^n,V^n=v^n|H=1) ~\mathbb{P}(\mathcal{D}_{0} |~U^n=u^n,V^n=v^n,~ \bar{\mathcal{E}}_0) \notag \\
=& \sum_{u^n,v^n}\mathbb{P}(U^n=u^n,V^n=v^n|H=1) ~ \sum_{s^n,y^n}\mathbb{P}(S^n=s^n,Y^n=y^n,\mathcal{D}_{0} |~U^n=u^n,V^n=v^n,~ \bar{\mathcal{E}}_0) \notag \\
=& \sum_{u^n,v^n}\mathbb{P}(U^n=u^n,V^n=v^n|H=1)~ \Big[~\sum_{s^n,y^n}\mathbb{P}(S^n=s^n,Y^n=y^n |~U^n=u^n,V^n=v^n,~ \bar{\mathcal{E}}_0) \notag \\
&\qquad \qquad \qquad \qquad \qquad \qquad \qquad  \qquad \qquad \mathbb{P}(\mathcal{D}_{0}|~U^n=u^n,V^n=v^n,S^n=s^n,Y^n=y^n,~ \bar{\mathcal{E}}_0) \Big] \notag \\
=&\sum_{u^n,v^n}\mathbb{P}(U^n=u^n,V^n=v^n|H=1) \Big[~\sum_{s^n,y^n}\mathbb{P}(S^n=s^n,Y^n=y^n |~ U^n=u^n, \bar{\mathcal{E}}_0) \notag \\
& \qquad \qquad \qquad \qquad \qquad \qquad \qquad  \qquad \qquad \mathbb{P}(\mathcal{D}_{0}|~U^n=u^n, ~V^n=v^n,S^n=s^n,Y^n=y^n,~ \bar{\mathcal{E}}_0) \Big]\label{spclfinsplt} && 
\end{flalign}
The event $\mathcal{D}_0$ occurs only if there exists a sequence $(\bar W^n(l),V^n,S^n,Y^n) \in T_{[\bar WVSY]_{\tilde \delta}}^n$ for some $l \in [e^{n(I(U,S;\bar W)+\mu)}]$. 
Noting that the quantization codebook is independent of the $(V^n,S^n,Y^n)$ given that $ \bar{\mathcal{E}}_0$ holds, it can be shown using standard arguments that
\begin{align}
\mathbb{P}(\mathcal{D}_{0}|~U^n=u^n,V^n=v^n,S^n=s^n,Y^n=y^n,~ \bar{\mathcal{E}}_0) \leq e^{-n(I(\bar W;V,Y|S)-I(U;\bar W|S)-O(\delta))}. \label{spt2bnd}
\end{align}
Also,
\begin{align}
\mathbb{P}(S^n=s^n,Y^n=y^n |~U^n=u^n, \bar{\mathcal{E}}_0) \leq e^{-n(H(\tilde S \tilde Y|\tilde U)+D(P_{\tilde S \tilde Y|\tilde U}||\check Q_{SY|U}|P_{\tilde U}))}. \label{probtypes}
\end{align}
Hence, using \eqref{spt2bnd} and \eqref{probtypes} in \eqref{spclfinsplt}, 
we obtain
 \begin{flalign}
& \sum_{u^n,v^n}\mathbb{P}(U^n=u^n,V^n=v^n|H=1) ~\mathbb{P}(\mathcal{D}_{0} |~U^n=u^n,V^n=v^n,~ \bar{\mathcal{E}}_0) \notag &&
\end{flalign}
\begin{flalign}
& \leq  (n+1)^{|\Ucal||\V||\Scal||\Y|}\max_{\substack{P_{\tilde U \tilde V \tilde S \tilde Y}: \\P_{\tilde V \tilde S \tilde Y}= \hat P_{VSY}}}e^{n H(\tilde U \tilde V \tilde S \tilde Y)} ~ e^{-n (H(\tilde U\tilde V)+D(P_{\tilde U\tilde V}||Q_{UV}))} ~   e^{-n(H(\tilde S \tilde Y|\tilde U)+D(P_{\tilde S \tilde Y|\tilde U}||\check Q_{SY|U}|P_{\tilde U}))} \notag \\ 
&\qquad \qquad \qquad\qquad\qquad \qquad\qquad e^{-n(I(\bar W;V,Y|S)-I(U;\bar W|S)-O(\delta))} \notag \\
&= e^{-nE_{3n}^*}, \notag &&
\end{flalign}
where,
\begin{align}
E_{3n}^*&=\min_{P_{\tilde V \tilde S \tilde Y}= \hat P_{VSY}}D(P_{\tilde V \tilde S \tilde Y}||\check Q_{VSY})+I(\bar W;V,Y|S)-I(U;\bar W|S)- |\Ucal||\V||\Scal||\Y| \log(n+1)-O(\delta) \notag \\ &\xrightarrow{(n)} E_3'\left(P_S,P_{\bar W|US},P_{X'|US},P_{X|US\bar W}\right)-O(\delta). \notag 
\end{align}
 Since the error-exponent is lower bounded  by the minimal value of the exponent due to the various type 2 error events, this completes the proof of the theorem.

 \section{Optimal single-letter characterization of error-exponent when $C(P_{Y|X})=0$} \label{proofzerocaperror-exponent}
The proof of achievability follows from the one-bit scheme mentioned in Remark \ref{remkzerratecompsch} which states that for $\tau \geq 0$, $\kappa(\tau, \epsilon) \geq \kappa_0(\tau)$, $\forall~\epsilon \in (0,1]$.
 Now, it is well-known (see \cite{Csiszar-Korner}) that $C(P_{Y|X})=0$ only if
 \begin{align}
  P^*_Y:= P_{Y|X=x}=P_{Y|X=x'},~\forall~x,x' \in \X. \label{allrowsequ}
 \end{align}
 From \eqref{allrowsequ}, it follows that $E_c(P_{Y|X})=0$. Also, 
 \begin{flalign}
     \beta_0 &\geq D(P_V||Q_V)+\min_{\substack{P_{\tilde U \tilde V}:\\ P_{\tilde U}=P_U,~P_{\tilde V}=P_V}}D(P_{\tilde U| \tilde V}||Q_{U|V} \big | P_{\tilde V}) \notag \\
     & \geq  D(P_V||Q_V), \notag &&
 \end{flalign}
which implies that $\kappa_0(\tau) \geq D(P_V||Q_V)$.

 \textit{Converse:}
 We first show the weak converse, i.e.,  $\kappa(\tau) \leq D(P_V||Q_V)$, where $\kappa(\tau)$ is as defined in \eqref{error-exponentforzerot1err}. For any sequence of encoding functions $f^{(k,n_k)}$ and acceptance regions $\mathcal{A}_{(k,n_k)}$ for $H_0$ that satisfy $n_k \leq \tau k$ and \eqref{vanisht1prconst}, it follows similarly to \eqref{divaserror-exponent}, that
 \begin{align}
     \limsup_{k \rightarrow \infty}  \frac{-1}{k}\log \left( \beta \left(k,n_k, f^{(k,n_k)}, g^{(k,n_k)} \right) \right) &  \leq \limsup_{k \rightarrow \infty} \frac 1k D\left(P_{Y^{n_k}V^k}|| Q_{Y^{n_k}V^k}\right). \label{rhsdiv} 
 \end{align}
The terms in the R.H.S. of \eqref{rhsdiv} can be expanded as
 \begin{flalign}
    & \frac 1k D\left(P_{Y^{n_k}V^k}|| Q_{Y^{n_k}V^k}\right) \notag \\ 
    &=D(P_V||Q_V)+\frac{1}{k} \sum_{\substack{(v^k,y^{n_k}) \\ \in \V^k \times \Y^{n_k}}} P_{V^kY^{n_k}}(v^k,y^{n_k}) \log \left( \frac{P_{Y^{n_k}|V^k}(y^{n_k}|v^k)}{Q_{Y^{n_k}|V^k}(y^{n_k}|v^k)} \right). \label{spliteq}&&
 \end{flalign}
 Next, note that
 
 \begin{flalign}
    P_{Y^{n_k}|V^k}(y^{n_k}|v^k)&= \sum_{\substack{(u^k,x^{n_k}) \\\in~ \Ucal^k \times \X^{n_k}}} P_{U^k|V^k}(u^k|v^k)P_{X^{n_k}|U^k}(x^{n_k}|u^k) P_{Y^{n_k}|X^{n_k}}(y^{n_k}|x^{n_k}) \notag \\
    &=\left( \prod_{i=1}^{n_k}P^*_{Y}(y_{i})\right) \sum_{\substack{(u^k,x^{n_k}) \\ \in~ \Ucal^k \times \X^{n_k}}} P_{U^k|V^k}(u^k|v^k)P_{X^{n_k}|U^k}(x^{n_k}|u^k) \label{allrowschnequal} \\
    &=\prod_{i=1}^{n_k}P^*_{Y}(y_{i}), \label{eqtorow1} &&
    \end{flalign}
 where,  \eqref{allrowschnequal} follows from \eqref{DMCprbdist} and \eqref{allrowsequ}. Similarly, it follows that
 \begin{align}
    Q_{Y^{n_k}|V^k}(y^{n_k}|v^k)&=\prod_{i=1}^{n_k}P^*_{Y}(y_{i}). \label{eqtorow2}
 \end{align}
From \eqref{rhsdiv}, \eqref{spliteq}, \eqref{eqtorow1} and \eqref{eqtorow2}, we obtain that
\begin{align}
     \limsup_{k \rightarrow \infty}  \frac{-1}{k}\log \left( \beta \left(k,n_k, f^{(k,n_k)}, g^{(k,n_k)} \right) \right) &  \leq D(P_V||Q_V). \notag 
\end{align}
This completes the proof of the weak converse. 

Next, we proceed to show that $D(P_{V}||Q_V)$ is the optimal error-exponent for every $\epsilon \in (0,1)$. For any fixed $\epsilon \in (0,1)$, let $f^{(k,n_k)}$ and  $\mathcal{A}_{(k,n_k)}$ denote any encoding function and acceptance region for $H_0$, respectively, such that $n_k \leq \tau k$ and 
\begin{equation} \label{t1errprobconstsingobs}
 \limsup_{k \rightarrow \infty} \alpha\left(k,n_k, f^{(k,n_k)}, g^{(k,n_k)} \right) \leq \epsilon.  
\end{equation}
The joint distribution of $(V^k,Y^{n_k})$ under the null and alternate hypothesis is given by
\begin{flalign}
    P_{V^kY^{n_k}}(v^k,y^{n_k})&= \left( \prod_{i=1}^k P_V(v_i)\right) \left(\prod_{j=1}^{n_k} P^*_Y(y_j) \right), \label{jntdistprodnull} \\
  \mbox{and }  Q_{V^kY^{n_k}}(v^k,y^{n_k})&=  \left( \prod_{i=1}^k Q_V(v_i)\right) \left(\prod_{j=1}^{n_k} P^*_Y(y_j) \right),   
\end{flalign}
respectively.
By the weak law of large numbers, for any $\delta>0$, \eqref{jntdistprodnull} implies that
\begin{align}
\lim_{k \rightarrow \infty} P_{V^kY^{n_k}}\left( T_{[P_V]_{\delta}}^k \times T_{[P^*_Y]_{\delta}}^{n_k} \right)&=1. \label{prodprobone}
\end{align}
Also, from \eqref{t1errprobconstsingobs}, we have
\begin{align}
 \liminf_{k \rightarrow \infty} P_{V^kY^{n_k}}\left(\mathcal{A}_{(k,n_k)} \right) \geq (1- \epsilon). \label{lbaccepregprb} 
\end{align}
From \eqref{prodprobone} and \eqref{lbaccepregprb}, it follows that
\begin{align}
  P_{V^kY^{n_k}} \left( \mathcal{A}_{(k,n_k)} \cap T_{[P_V]_{\delta}}^k \times T_{[P^*_Y]_{\delta}}^{n_k} \right) \geq 1-\epsilon', \label{typicsetovrlap}
\end{align}
for any $\epsilon' > \epsilon$ and $k$ sufficiently large ($k \geq k_0(\delta,|\V|,|\Y|)$). Let
\begin{align}
\mathcal{A}(v^k, \delta)&:= \left\lbrace y^{n_k}: (v^k,y^{n_k}) \in \mathcal{A}_{(k,n_k)} \cap T_{[P_V]_{\delta}}^k \times T_{[P^*_Y]_{\delta}}^{n_k}\right\rbrace,\\
\mbox{and }\mathcal{D}(\eta, \delta)&:= \left\lbrace v^k \in T_{[P_V]_{\delta}}^k: P_{Y^{n_k}}(\mathcal{A}(v^k,\delta)) \geq \eta \right\rbrace. \label{detasetdef}
\end{align}
Fix $0<\eta'< 1-\epsilon'$. Then, we have from \eqref{typicsetovrlap} that for any $\delta>0$ and sufficiently large $k$,
\begin{align}
    P_{V^k} \left( \mathcal{D}(\eta', \delta)\right) \geq \frac{1- \epsilon'-\eta'}{1-\eta'}. \label{lbforsideinfseq}
\end{align}
From \cite[Lemma 2.14]{Csiszar-Korner}, \eqref{lbforsideinfseq} implies that $\mathcal{D}(\eta', \delta)$ should contain atleast $\frac{1- \epsilon'-\eta'}{1-\eta'}$ fraction  (approx.) of sequences in $T_{[P_V]_{\delta}}^k$ and for each $v^k \in \mathcal{D}(\eta', \delta)$, \eqref{detasetdef} implies that $\mathcal{A}(v^k, \delta)$ should contain atleast $\eta'$ fraction (approx.) of sequences in $T_{[P^*_Y]_{\delta}}^{n_k}$, asymptotically. Hence, for sufficiently large $k$, we have
\begin{align}
 Q_{V^kY^{n_k}}\left(\mathcal{A}_{(k,n_k)}\right) &\geq  \sum_{v^k \in \mathcal{D}(\eta', \delta)} Q_{V^k}(v^k) \sum_{y^{n_k} \in \mathcal{A}(v^k, \delta)} P_{Y^n}(y^{n_k}) \\
& \geq e^{-k \left( D(P_V||Q_V)- \frac{\log\left(\frac{1- \epsilon'-\eta'}{1-\eta'}\right)}{k}- \frac{\log(\eta')}{k}-O(\delta)\right)}. \label{oldt2e}
\end{align}
Here, \eqref{oldt2e} follows from \cite[Lemma 2.6]{Csiszar-Korner}.

Let $\mathcal{A}'_{(k,n_k)}:=T_{[P_V]_{\delta}}^k \times T_{[P^*_Y]_{\delta}}^{n_k}$. Then, for sufficiently large $k$, 
\begin{align}
    P_{V^kY^{n_k}}\left(\mathcal{A}'_{(k,n_k)}\right) &\xrightarrow{(k)} 1, \label{probtendsto1}
    \\  \mbox{ and } Q_{V^kY^{n_k}}\left(\mathcal{A}'_{(k,n_k)}\right)  &\leq  e^{-k \left( D(P_V||Q_V)-O(\delta)\right)}, \label{newt2ezerot1}
\end{align}
where, \eqref{probtendsto1} and \eqref{newt2ezerot1} follows from weak law of large numbers and \cite[Lemma 2.6]{Csiszar-Korner}, respectively.
Together \eqref{oldt2e}, \eqref{probtendsto1} and \eqref{newt2ezerot1} implies that 
\begin{align}
    |\kappa(\tau, \epsilon)-\kappa(\tau)| \leq O(\delta), \notag 
\end{align}
and the proposition is proved  since $\delta>0$ is arbitrary.

\end{appendices}

\bibliographystyle{IEEEtran}
\bibliography{refarxiv}

\end{document}